\documentclass[11pt,preprintnumbers,pra,aps,
nofootinbib,tightenlines,floatfix, twocolumn,superscriptaddress,longbibliography]{revtex4-2}

\usepackage[colorlinks=true]{hyperref}
\hypersetup{
  linkcolor  = TealBlue!90!Black,
  citecolor  = YellowGreen!90!Black,
  urlcolor   = Purple!90!Black,
  colorlinks = true,
}
\usepackage[dvipsnames]{xcolor}
\usepackage[all]{hypcap}
\usepackage{amsmath,amssymb}
\usepackage{graphicx}
\usepackage{physics}
\usepackage{mathtools}
\usepackage{balance}
\usepackage{cancel}
\usepackage[utf8]{inputenc}
\usepackage{csquotes}
\usepackage{empheq}
\usepackage{longtable}
\usepackage{multirow}
\usepackage[section]{placeins}
\usepackage{comment}
\usepackage[customcolors]{hf-tikz}
\usepackage[top=1in,bottom=1in,left=1.8cm,right=1.6cm]{geometry}

\DeclareMathOperator{\diag}{diag}

\allowdisplaybreaks
\makeatletter
\renewcommand\onecolumngrid{
\do@columngrid{one}{\@ne}
\def\set@footnotewidth{\onecolumngrid}
\def\footnoterule{\kern-6pt\hrule width 1.5in\kern6pt}%
}
\makeatother

\newcommand{\dif}{\ensuremath{\text{d}}}
\newcommand\scalemath[2]{\scalebox{#1}{\mbox{\ensuremath{\displaystyle #2}}}}

\catcode`,\active

\catcode`\,12

\def\Nslash{{ {\cal N}\hskip-0.55em /}}
\def\phipislash{\cancelto{}{\phi\pi}}

\newcount\hour \newcount\hourminute \newcount\minute
\hour=\time \divide \hour by 60
\hourminute=\hour \multiply \hourminute by 60
\minute=\time \advance \minute by -\hourminute

\DeclareMathOperator{\nullspace}{null}

\begin{document}

\title{Finite Gaussian assistance protocols and a conic metric for extremizing \texorpdfstring{\\}{} spacelike vacuum entanglement}

\author{Boyu Gao}
\email{boyu.gao@duke.edu}
\affiliation{{Duke Quantum Center and Department of Physics, Duke University, Durham, NC 27708, USA}}
\author{Natalie Klco}
\email{natalie.klco@duke.edu}
\affiliation{{Duke Quantum Center and Department of Physics, Duke University, Durham, NC 27708, USA}}

\begin{abstract}

In a pure Gaussian tripartition, a range of entanglement between two parties ($AB$) can be purified through classical communication of Gaussian measurements performed within the third ($C$). To begin, this work introduces a direct method to calculate a hierarchic series of projective $C$ measurements for the removal of any $AB$ Gaussian noise, circumventing divergences in prior protocols. Next, a multimode conic framework is developed for pursuing the maximum (Gaussian entanglement of assistance, GEOA) or minimum (Gaussian entanglement of formation, GEOF) pure entanglement that may be revealed or required between $AB$. Within this framework, a geometric necessary and sufficient entanglement condition emerges as a doubly-enclosed conic volume, defining a novel distance metric for conic optimization. Extremizing this distance for spacelike vacuum entanglement in the massless and massive free scalar fields yields (1) the highest known lower bound to GEOA, the first that decays slower than the two-point correlation functions and (2) the lowest known upper bound to GEOF, the first that decays exponentially mirroring the mixed $AB$ negativity. Furthermore, combination of the above with a generalization of previous partially-transposed noise filtering techniques allows calculation of a single $C$ measurement that maximizes the purified $AB$ entanglement. Beyond expectation that these behaviors of spacelike GEOA and GEOF persist in interacting theories, the present measurement and optimization techniques are applicable to physical many-body Gaussian states beyond quantum fields.

\end{abstract}

\date{\today}
\maketitle

{
\small
\tableofcontents
}

\section{Introduction}
\vspace{0.2cm}

Many-body entanglement is a complex resource that is pervasive in the fundamental fields of nature~\cite{ReehSchlieder,SUMMERS1985257,summers1987maximal,summers1987bell1,summers1987bell2,Witten} and key to the operating systems of quantum technologies. Understanding and engaging with such large-scale quantum correlations continues to be crucial for progress~\cite{preskillarticle}, from dualities toward a quantum description of gravity~\cite{Maldacena:1997re,ShinseiRyu_2006} to catalyzing topological features in error-robust quantum computations~\cite{Shor:1995hbe,Calderbank:1995dw,phdthesisgottesman1997stabilizer}. Naturally, this complexity is challenging to express classically, especially for large mixed quantum states such as those observed by spacelike field detectors. However, the Gaussian formalism has allowed first calculations of spacelike entanglement to be convergent into the infinite-body regime of continuum quantum fields~\cite{scalar1dextra,NKnegativitysphere,gao2024detectingspacelikevacuumentanglement}. This paper creates new techniques for optimizing pure entanglement resources underlying mixed quantum states, extending the reach of Gaussian quantum information to previously intractable measures of many-body entanglement.

\begin{figure}[t!]
\centering
\includegraphics[width=0.5\textwidth]{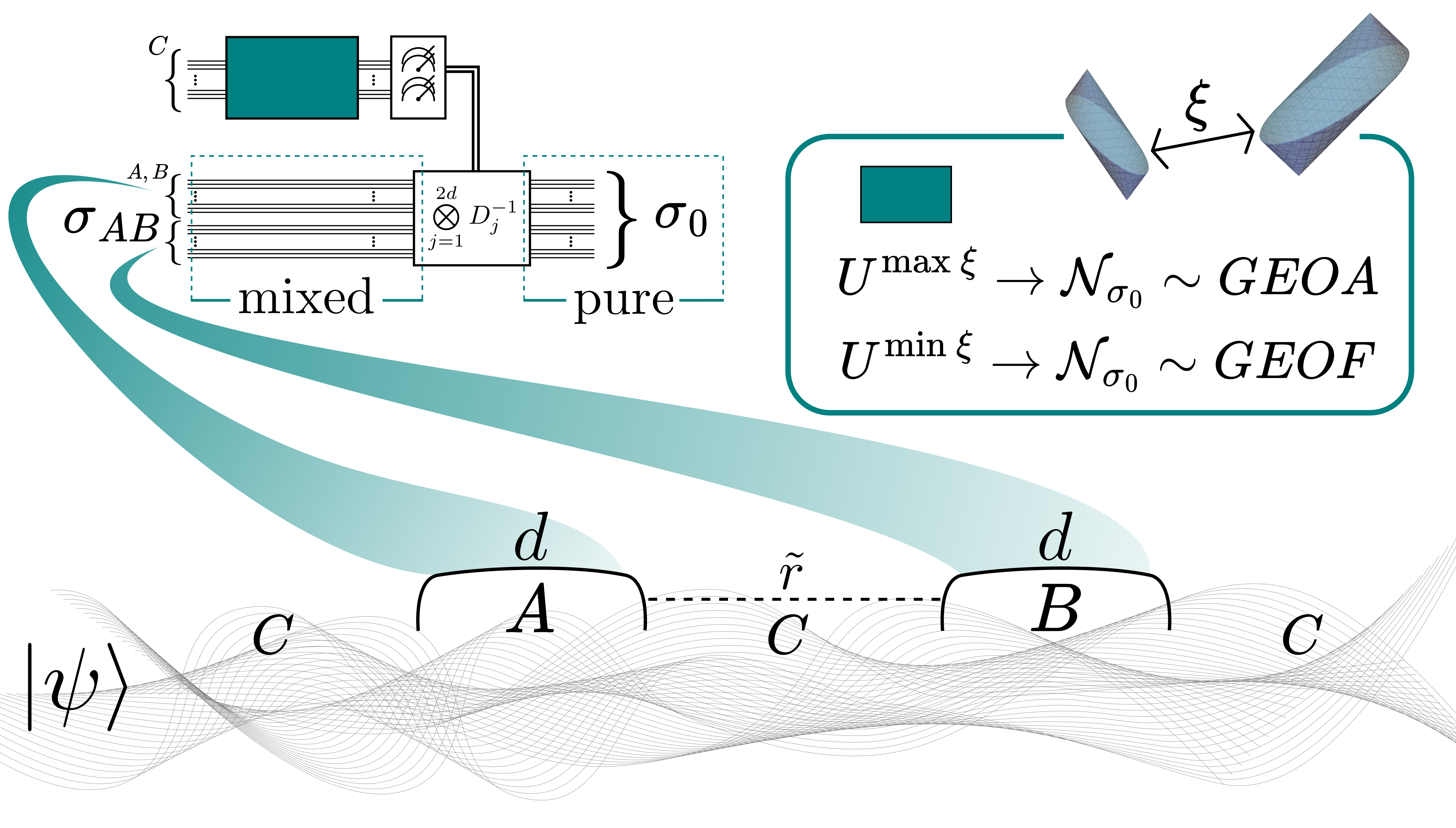}
\caption{Diagram illustrating the assisted entanglement protocol applied to the scalar field vacuum. The global vacuum is partitioned into three spatial regions: two disjoint patches, $A$ and $B$, each of size $d$ and separated by distance $\tilde{r}$, with the external volume $C$. As depicted in the quantum circuit, classical communication of measurements within $C$ informs controlled single-mode displacements that identify pure entanglement resources shared between $A$ and $B$. The teal unitary gate defines the (entangled) measurement basis followed by single-mode measurements along the quadrature axes, $\phi$ or $\pi$. The present work calculates extremal entanglement purified in $AB$ corresponding to a tight lower(upper) bound on GEOA(GEOF), achieved by derivations of $U$ maximizing(minimizing) the multimode conic distance metric $\xi$ introduced in Sec.~\ref{sec:IIIA}.}
\label{fig:diagram}
\end{figure}

Due to the nontrivial role of classical correlations in quantum information processing, a multitude of observables are needed to quantitatively characterize mixed-state entanglement. For example, the logarithmic negativity~\cite{peresoriginalN,HORODECKIoriginalN,Simonreflection,computablemeasure,PlenioLogarithmic} between disjoint regions of scalar field vacuum~\cite{scalar1dextra,Calabrese:2012ew,NKnegativitysphere,NKentsphere} quantifies the amount of entanglement that may be extracted by pairs of locally interacting detectors~\cite{NKcorehalo,gao2024detectingspacelikevacuumentanglement}, while the vacuum correlation functions follow the (exponentially larger) amount of entanglement that may be purified via communication of spatially local external measurements~\cite{NKvolumemeasure}. This work focuses on two additional operational properties of vacuum quantum correlations: (1) the Gaussian entanglement of assistance (GEOA), a tripartite entanglement measure maximizing the pure resource that may be concentrated in $AB$ via communication of arbitrary measurements in $C$~\footnote{Assisted entanglement~\cite{PhysRevLett.80.2493,divincenzo1999entanglement,laustsen2002local,Smolin_2005,PhysRevA.73.062331,PhysRevA.101.052305} in Gaussian contexts has been shown to be equal to Entanglement of Collaboration~\cite{PhysRevA.101.052305}, the generalized quantity allowing two-way communication~\cite{PhysRevA.73.062331}.  Given the equivalence, this work adopts the simpler language of the GEOA formulation.} and (2) the Gaussian entanglement of formation (GEOF), a bipartite entanglement measure minimizing the amount of pure-state entanglement required to prepare the $AB$ mixed state. Interestingly, both GEOA and GEOF can be formulated within the former assisted paradigm, depicted in Fig.~\ref{fig:diagram}, due to a fundamental correspondence between external Gaussian measurements and classical noise decompositions~\cite{PhysRevA.101.052305,8004445} that is further strengthened in Sec.~\ref{sec:IIA}. With communication of entangled $C$ measurements capable of affecting the pure $AB$ quantum resources, we will see that neither of these operational measures of entanglement scale simply as the two-point correlation functions of the field.

The culminating results of this work are calculations of explicit underlying pure states that provide lower(upper) bounds on the GEOA(GEOF), which exhibit scaling behaviors with vacuum separation that are parametrically tighter than prior estimates~\cite{NKvolumemeasure,gao2024partialtransposeguided}. Specifically, the lower bound on GEOA for massive(massless) fields achieves an asymptotically constant(double-logarithmically decaying) entanglement at long distances, reminiscent of the AKLT state~\cite{Affleck:1987vf}, while the upper bound on GEOF achieves an exponential decay mirroring the region-region logarithmic negativity. As such, it is shown that measurements on the external field volume ($C$) can generate a wide range of pure $AB$ entanglement scalings---from constant to logarithmic, polynomial, and exponential. The finite measurement-noise correspondence in this work further derives hierarchies of measurements capable of producing each of these decays.

The presented calculations are enabled by introducing a new geometric framework of double-cone volumes ($DCV$s) for multimode Gaussian systems, with an inter-$DCV$ distance metric $\xi$ that serves as a necessary and sufficient entanglement quantifier. The creation of this metric, which guides optimizations throughout this work, is made possible by a detailed understanding of the semidefinite cone structure associated with Gaussian classical mixing channels.

This article is organized as follows. After presenting an overview of Gaussian states at the beginning of Section~\ref{sec:II}, Section~\ref{sec:IIA} discusses a concrete protocol for calculating projective measurements corresponding to Gaussian noise decompositions, with details elaborated in Appendix~\ref{sec:appa} and demonstrated for a scalar field vacuum in Appendix~\ref{sec:appbsec3} and~\ref{sec:appcsec4}. Section~\ref{sec:IIB} discusses the optimization of underlying pure-state entanglement for two-mode Gaussian states within the $DCV$ framework, with complete derivations of the maximization and minimization provided in Appendices~\ref{sec:appbsec1} and~\ref{sec:appcsec1}. Creating a $DCV$ framework for multimode Gaussian states, the inter-$DCV$ distance metric is introduced in Section~\ref{sec:IIIA}, with geometric foundations provided in Appendix~\ref{sec:appcsec2}. Finally, calculations bounding the GEOA and GEOF by maximizing and minimizing this distance are shown in Section~\ref{sec:IIIB}, with full derivations formulated in Appendices~\ref{sec:appcsec3} and~\ref{sec:appcsec4}, respectively.

\section{Entanglement in Gaussian states}
\label{sec:II}

Defining the vector of phase-space operators~\cite{It1,It3,BraunsteinGaussiareview,It2,horodecki2009quantum,WeedbrookGaussiareview,serafini2017quantum} as $\hat{\boldsymbol{r}} \equiv \left( \hat{\phi}_1,...\hat{\phi}_n,\hat{\pi}_1,...,\hat{\pi}_n\right)^{T}$ the canonical commutation relations (CCRs) take the form, ${\left[ \hat{\boldsymbol{r}}, \hat{\boldsymbol{r}}^{T} \right] = i\Omega}$. The symplectic matrix $\Omega$ is given by 
\begin{equation} 
\Omega = \begin{pmatrix} 0 & \mathbb{I} \\ -\mathbb{I} & 0 \\ \end{pmatrix} \ \ \ ,
\end{equation}
where $\mathbb{I}$ is an $n$-dimensional identity matrix. Any valid symplectic transformation, $S$, acting on the $n$ pairs of position and momentum operators must preserve the CCRs, $S \Omega S^{T} = \Omega$. In the application to spacelike field entanglement, the number of modes in $A$ and $B$ will be equal, $d = N_{AB}/2$, and their ordering in $\mathbf{r}$ will reverse A from lattice order, as in Ref.~\cite{gao2024detectingspacelikevacuumentanglement}, to manifest symmetry under the interchange of the $A$ and $B$ local Hilbert spaces.

Gaussian states are quantum states that are Gaussian in phase space, which may be completely characterized by the vector of first moments $\bar{\boldsymbol{r}}$ and Covariance Matrix (CM), 
\begin{equation} \sigma=\mathrm{Tr} \left[ \rho_{G} \left\{ (\hat{\boldsymbol{r}} - \bar{\boldsymbol{r}}),(\hat{\boldsymbol{r}} - \bar{\boldsymbol{r}})^{T} \right\} \right] \ \ \ , 
\end{equation}
where $\rho_G$ is the density matrix. With local displacement operators that do not affect entanglement, the first moments can be set to zero. As such, for Gaussian states, all quantum and classical correlations between modes are encoded in the CM. Any physical CM can be transformed into its Williamson normal form~\cite{williamsonnormalform} through a global symplectic transformation, $S \sigma S^{T} = D \oplus D$, where $D \geq \mathbb{I}$ is a diagonal matrix containing the real, positive symplectic eigenvalues $\nu_j \in \text{spec} | i \Omega \sigma| $ along its diagonal.

According to the Peres-Horodecki criterion~\cite{peresoriginalN,HORODECKIoriginalN,Simonreflection}, a sufficient condition for entanglement in quantum systems with any number of modes is the presence of a negative eigenvalue in the partially transposed (PT) density matrix. For Gaussian states, this entanglement criterion is associated with the physicality of the PT CM, denoted as $\Tilde{\sigma} = \Lambda \sigma \Lambda$, where $\Lambda_{d}=(\mathbb{I}_{d} \oplus \mathbb{I}_{d} \oplus \mathbb{I}_{d} \oplus -\mathbb{I}_{d})$ is the momentum reversal operator within the $B$ half-space~\cite{Simonreflection}. Logarithic negativity~\cite{computablemeasure, PlenioLogarithmic} is an entanglement monotone created from this PT space in the form, 
\begin{equation}
    \mathcal{N} = -\sum_{j=1}^{n_{-}}\log_{2}\Tilde{\nu}_{j} \ \ \ ,
    \label{eq:negg}
\end{equation}
where $n_- \leq d$ denotes the number of PT symplectic eigenvalues, $\tilde{\nu}_j$, that are smaller than one. These $\tilde{\nu}_j$ can be determined from $\text{spec} | i \Omega \tilde{\sigma}| $ or a global symplectic transformation, $\Tilde{S}$, that brings the PT CM to normal form, $\Tilde{S} \Tilde{\sigma} \Tilde{S}^{T} = \Tilde{D} \oplus \Tilde{D}$, where $\Tilde{D} \geq 0$ is a diagonal matrix containing the real, positive PT symplectic eigenvalues along its diagonal. For disjoint regions of the free scalar field vacuum, the logarithmic negativity is UV-finite~\cite{scalar1dextra} and provides a necessary and sufficient condition for separability~\cite{gao2024detectingspacelikevacuumentanglement}.

Pure Gaussian CMs can be parametrized in block form~\cite{WolfGEOF},
\begin{equation}
\sigma_{\text{pure}}  = \begin{pmatrix}
X & XY \\
YX & YXY+X^{-1} \\
\end{pmatrix} \ \ \ ,
\label{eq:pureCMform}
\end{equation}
with $X>0$ and symmetric $X$, $Y$. The corresponding normalized pure Gaussian wave function with vanishing first moments is,
\begin{equation}
\scalemath{1}{|\psi\rangle = \frac{\det  X^{-\frac{1}{4}}}{\pi^{\frac{N}{4}}} \int \dif \phi \ e^{-\frac{1}{2} \phi^T \left( X^{-1} - iY \right) \phi}  \ |\phi\rangle }  \ \ \ .
\label{eq:vacuumWF}
\end{equation}
This parametrization (Eqs.~\eqref{eq:pureCMform} and~\eqref{eq:vacuumWF}) will be employed throughout the present paper, offering analytic and numerical advantages in calculating multimode entanglement properties involving a global pure state or underlying pure state convex decompositions.

\subsection{Associating projective measurements to Gaussian noise decompositions}
\label{sec:IIA}

For computational purposes, it can be advantageous to view the GEOA from an alternate perspective of extremizing with respect to viable $AB$ convex decompositions of a pure state subject to the Gaussian classical mixing channel~\cite{PhysRevA.101.052305}. Thus, the measurement protocol of Fig.~\ref{fig:diagram} can be regarded as a dilated procedure for removing Gaussian noise $Y = \sigma_{AB}-\sigma_0$ with the assistance of entangled measurements in the $C$ space of an initial $ABC$ purification of $\sigma_{AB}$. In this section, we formalize this idea by introducing a measurement protocol in $C$ that may be determined to achieve any such noise removal. Our strategy performs measurements in the (evolving) basis of local normal modes, repeatedly simplifying the global pure state into a tensor product of two-mode squeezed vacuum  states (TMSVSs)~\cite{PhysRevA.67.052311, Giedkepurestatetrans} spanning the $AB|C$ partition. Furthermore, by working directly in terms of projective measurements, our protocol circumvents divergences that arise in formulations with Gaussian Positive Operator Valued Measures (POVMs)~\cite{8004445,PhysRevA.101.052305}. The following discussion focuses on the conceptual framework of this connection between measurements and quantum noise, while Appendix~\ref{sec:appa} provides the complete protocol for readers interested in the technical details.

It is well known that implementing a general measurement can be performed via unitary dynamics, projective measurements, and an ancilla system~\cite{Nielsen_Chuang_2010}. As depicted in the circuit of Fig.~\ref{fig:diagram}, projective $C$-measurements in entangled bases may be performed via a preceding unitary transformation localizing collective modes of interest to each site, followed by tensor-product projections into the $\phi$ or $\pi$ quadrature of each mode. Following procedures in Ref.~\cite{NKvolumemeasure} for $\phi\pi$-uncorrelated pure CMs or Appendix~\ref{sec:appasec2} for $\phi\pi$-correlated pure CMs, it is seen that the resulting pure $AB$ CM is dependent on the measurement basis, but independent of the measurement result~\cite{gaussianoperation,NKvolumemeasure,8004445,PhysRevA.101.052305,Fiurasek:2002tvf}. Hence, classical communication informing measurement-controlled displacements is sufficient to homogenize all pure states from an ensemble, eliminating the noise and allowing focus to proceed in the context of a single CM.

Given an $AB$ noise decomposition and an $ABC$ purification, this work details a constructive procedure for associating each component of a rank-one decomposition of $Y$ with a collective operator profile for a single projective $C$ measurement (Appendix~\ref{sec:appasec4}). The strategy presented involves computing each profile sequentially in the basis of $AB$ and $C$ local normal modes, which evolves after each measurement. In calculation, each measurement is performed after symplectic transformations that place the profile of phase space operators onto one mode, which is equivalent in practice to implementing a single symplectic transformation in C followed by projectively measuring the last rank(Y) modes (e.g., illustrated in Appendix Fig.~\ref{fig:circuitdiagram}). If the number of purities ($\nu = 1$) in the (non-degenerate) symplectic spectrum of the $AB$ CM, $P_{\sigma_{AB}}$, increases upon subtraction of a rank-one noise component, an auxiliary mode is not required for the associated projective measurement (Appendix~\ref{sec:appasec3}). However, if this condition is not satisfied, use of an auxiliary mode becomes necessary. In general, the number of required auxiliary modes is shown to be upper bounded by $2N_{AB} - P_{\sigma_{AB}} - P_{\sigma_{0}}$ (Appendix~\ref{sec:appasec4}). Thus, for pure $\sigma_0$ and zero purities in $\sigma_{AB}$, the number of auxiliary modes is upper bounded by the number of modes in $AB$, $N_{AB}$. 

When the number of bosonic modes in $C$ is greater than that in $AB$, as is the case in the application to disjoint regions of scalar field vacuum where the excess diverges in the infinite volume limit, there are modes in $C$ that may naturally be treated as auxiliary. In the local normal mode basis, the party with a greater number of modes will contain pure normal modes that are decoupled from the rest of the system. These decoupled modes can be utilized as the auxiliary modes for the measurement procedure. Thus, when the number of modes in $C$ is twice the number of modes in $AB$, no extra modes are required to perform measurements associated with the general noise decomposition. Note that extracting the post-measurement pure component from $AB$, or the corresponding collective mode for projective measurement in $C$, can be achieved through the beamsplitter interaction as outlined in Ref.~\cite{gao2024detectingspacelikevacuumentanglement}.

Finally, because rank-one decompositions are not unique, the relationship between measurement procedures and a Gaussian noise decomposition are many-to-one. This raises an opportunity to embed advantageous structure within the sequence of measurements. For example, through a generalization of the Minimum Noise Filtering (MNF) process of Ref.~\cite{gao2024partialtransposeguided} to an alternate PT subspace, Appendix~\ref{sec:appbsec3} demonstrates an exponential hierarchy of entanglement purified in the $AB$ scalar field vacuum regions that maximizes the production from the first $C$ measurement.

Previous works have viewed Gaussian noise decomposition protocols from the perspective of Gaussian POVMs in $C$ under the assisted paradigm (see proposition 4 in Ref.~\cite{PhysRevA.101.052305} and proposition 4 in Ref.~\cite{8004445}). For Gaussian noise that is full rank, the seeds for Gaussian POVMs in $C$ can be calculated directly~\cite{8004445}, and implemented via a symplectic transformation in $C$ followed by heterodyne detections for each POVM-mode~\footnote{For POVM-modes with finite single-mode squeezing, heterodyne detections are carried out using projective measurements, with one auxiliary bosonic mode introduced for each detection~\cite{PhysRevA.84.012119,genoni2014general,serafini2017quantum}.}. However, when the Gaussian noise is rank-deficient---as often occurs in the context of underlying pure state entanglement optimization (see, e.g., Ref.~\cite{WolfGEOF} and results from this work)---the corresponding seeds for the Gaussian POVMs exhibit divergences that hinder the direct normal mode decomposition. These divergences arise from infinite single-mode squeezing in certain POVM-modes. To find the corresponding modes via Ref.~\cite{8004445}, the appropriate limit involving products of matrices with increasingly small matrix power must be taken, where it is challenging to demonstrate convergence either numerically or analytically, especially in the multimode context. By replacing the Gaussian POVM formalism with an equally universal set of explicit physical resources---auxiliary modes, unitary transformations generated by bilinear Hamiltonians, and projective measurements in a quadrature basis---the current work allows all infinities to be handled naturally by the homodyne detection protocol~\footnote{The number of auxiliary modes required in the Gaussian POVMs formalism agrees with that identified in this work.}. As a result, this work provides a formalism that circumvents divergences, significantly reduces computational overhead, and provides a stable framework for calculation of assisted measurement protocols.

Given this correspondence, both operational entanglement measures of present interest can be framed in the language of extremizing the pure-state entanglement within $AB$ Gaussian classical mixing channels, 
\begin{equation}
  \mathcal{E}(\sigma) \equiv  \{ E(\sigma_{0}) \quad | \quad \forall \sigma_{0} \leq \sigma\} \ \ \ ,
\label{eq:keyeq}
\end{equation}
where $\sigma_0 \geq i \Omega$, $E(\cdot)$ denotes a valid entanglement measure~\footnote{While GEOA and GEOF are originally studied using the von Neumann entanglement entropy of the underlying pure state~\cite{divincenzo1999entanglement,mixstateent,WolfGEOF}, this work utilizes logarithmic negativity~\cite{computablemeasure,PlenioLogarithmic} due to its advantageous structure. The two are observed to exhibit the same scaling behavior.}, and $\mathcal{E}\left(\sigma_{AB}\right)$ is the set of possible entanglement between $AB$ upon noise removal. The pure case $\det \sigma_0 = 1$ is of particular interest for the entanglement analyses of this work.  However, present techniques organized by rank-one decompositions enable the correspondence to remain constructive for non-pure $\sigma_0$ (e.g., truncated as in Appendix~\ref{sec:appbsec3}).

This connection mirrors for GEOA that of Ref.~\cite{WolfGEOF}, which simplified the GEOF calculation into the Gaussian classical mixing channel by requiring that mixed Gaussian states have support only on components with faster-decaying correlations, i.e., $\sigma_0 \leq \sigma$. Thus, GEOF and GEOA can be viewed together as the two extremes of the assisted paradigm, highlighting the significance of both the maximization and minimization of the underlying pure-state entanglement. Due to the excessive degrees of freedom complicating operations in the $C$ space of present scalar field applications (e.g., spanning infinite modes in the result of Fig.~\ref{fig:plot} or demonstrated for finite field volume in Appendix~\ref{sec:appbsec3}), calculations of GEOA and GEOF are chosen to be carried out from the dual perspective of noise decompositions in the $AB$ space via extremizing Eq.~\eqref{eq:keyeq}. From this perspective arises a geometric framework of semidefinite cones.

\subsection{Quantifying entanglement with semidefinite cones: the two-mode \texorpdfstring{$DCV$}{DCV}}
\label{sec:IIB}

Any symmetric real two-dimensional matrix can be parametrized with Pauli matrices $\left( \mathbb{I}, X, Z\right)$, where the PSD condition defines a conic geometry in the associated three-dimensional parameter space. In Ref.~\cite{WolfGEOF}, this semidefinite cone framework is used in GEOF calculations for two-mode Gaussian states in normal form ($\sigma = \sigma_\phi \oplus \sigma_\pi$)~\cite{Simonreflection,Duanlocaltrans},
\begin{equation}
    \sigma_\phi \geq X \geq \sigma_\pi^{-1} \ \ \ ,
\end{equation}
such that the optimal pure Gaussian state (parametrized by $X$ of Eq.~\eqref{eq:pureCMform}) lives inside both the down- and up-ward semidefinite cone with apex located at $\sigma_{\phi}$ and $\sigma_{\pi}^{-1}$, respectively. Such a geometric space will hence be referred to as a Double-Cone Volume ($DCV$), i.e., the enclosed region formed when one cone in a double-cone structure is translated so that its apex lies within the other cone (e.g., illustrated in Appendix Fig~\ref{fig:dcvGeo_xpyplane}). It is further shown in Ref.~\cite{WolfGEOF} (with completed justification in Appendix~\ref{sec:appcsec1}) that optimal $X$ resides on the $DCV$-rim. By updating previous rim parametrizations and analytically extremizing the local purity, the derivations of Appendix~\ref{sec:appcsec1} provide a straightforward technique for fast and high-precision evaluation of optimal solutions along the rim, e.g., eliminating the role of coupled equations and numerical searching~\cite{ivan2008entanglementformationgaussianstates,marian2008entanglement,akbari2015entanglement,2019geof}.

In Ref.~\cite{PhysRevA.101.052305}, the calculation of GEOA was considered in the context of completely uncorrelated $\sigma = \sigma_A \oplus \sigma_B$ two-mode Gaussian states and Gaussian states with minimum negativity for fixed global and local purities (GLEMS)~\cite{adesso2004determination,It1}. In each of these cases, the optimal underlying pure state was found to live on the $DCV$-rim. In this paper, it is found that GEOA for symmetric two-mode Gaussian states can be calculated analytically. By invoking Weyl's inequality~\cite{hornmatrixana} and its generalization to symplectic eigenvalues~\cite{Giedkepurestatetrans}, the optimal solution can be constructed through the eigendecomposition of $\sigma_\phi-\sigma_\pi^{-1}$ (see Appendix~\ref{sec:appbsec1} for details), and is found to also live on the $DCV$-rim (i.e., Appendix Fig.~\ref{fig:coneconediagram1}). This solution exhibits clear duality with that of GEOF, each residing at one end of the semi-major axis of the elliptical rim~\footnote{Entanglement of formation (EOF) for an arbitrary two-mode Gaussian state has been shown to be equivalent to GEOF~\cite{akbari2015entanglement}, confirming the conjecture of Ref.~\cite{ivan2008entanglementformationgaussianstates} and pursued in Ref.~\cite{marian2008entanglement}. In contrast, the GEOA has been shown to differ from entanglement of assistance (EOA) in the three-mode case~\cite{PhysRevA.75.060302}, indicating that such duality does not generally extend beyond the Gaussian regime.}. The creation of a multimode extension of this framework is central to the achievements of Section~\ref{sec:III}.

\section{Entanglement in free scalar field vacuum}
\label{sec:III}

The free lattice scalar field vacuum is a Gaussian state, with its entanglement fully characterized by the CM of vacuum two-point correlation functions, $\langle 0 | \phi_i \phi_j | 0 \rangle$ and $\langle 0 | \pi_i \pi_j | 0 \rangle$, which may be analytically expressed with hypergeometric functions in the infinite volume limit. Despite the disjoint nature of spacelike field regions $A$ and $B$, the entanglement between the two does not vanish, as long as lattice artifacts at large separations are appropriately mitigated~\cite{NKnegativitysphere,NKentsphere,gao2024detectingspacelikevacuumentanglement}. 

Guiding the $(1_A \times 1_B)$-mode extractable entanglement in the vacuum~\cite{NKcorehalo,gao2024partialtransposeguided,gao2024detectingspacelikevacuumentanglement}, logarithmic negativity exhibits exponential decay with separation (linear(qudratic) exponent in the massless(massive) field)~\cite{scalar1dextra,NKnegativitysphere,NKentsphere}. This is distinct from the complementary observation that unentangled $C$-measurements ($U = \mathbb{I}$ in Fig.~\ref{fig:diagram}) yield pure resources $\mathcal{N}_{\sigma_0}^{\phi, \pi}$ that follow the two-point functions~\cite{NKvolumemeasure} (algebraic(linear-exponential) decay in the massless(massive) field) as shown for reference in Fig.~\ref{fig:plot}. Synthesizing the physical implications of these two decays has inspired study of the two-region GEOF, leading to a conjecture that it also decays exponentially~\cite{gao2024partialtransposeguided}. 

Within a new system of semidefinite cones discussed in Section~\ref{sec:IIIA}, this conjecture is numerically verified in Section~\ref{sec:IIIB}, where parametrically stronger bounds are presented for the maximization and minimization of underlying pure state entanglement.

\subsection{Multimode inter-DCV distance metric}
\label{sec:IIIA}

This section creates an alternate $DCV$ framework for multimode Gaussian states that are $AB$-symmetric and $\phi\pi$-uncorrelated, e.g., as are spacelike regions of the scalar field vacuum. For such states, the general necessary and sufficient separability condition framed as the existence of $\sigma_A, \sigma_B$ such that $\sigma \geq \sigma_A \oplus \sigma_B$~\cite{Gaussianboundent} may be reformulated as the existence of $X_A$ and $X_{AB} = 0$ such that,
\begin{equation}
    \sigma_{\phi,A} \pm \sigma_{\phi,AB} \geq X_{A}\pm X_{AB} \geq  \left(\sigma_{\pi}^{-1}\right)_{A} \pm \left(\sigma_{\pi}^{-1}\right)_{AB}  
    \label{eq:twodcv}
\end{equation}
(see Appendix~\ref{sec:appcsec2} for further details). The two expressions ($\pm$) of Eq.~\eqref{eq:twodcv} define two new $DCV$s, $DCV_{+}$ and $DCV_{-}$. The intersection of these multimode $DCV$s, i.e., a non-zero volume of $DCV_- \cap DCV_+$, is thus a necessary and sufficient separability condition in this context. The inter-$DCV$ separation, chosen to be quantified by the Frobenius norm $\xi = \|X_{AB}\| $, thus characterizes entanglement. With the definiteness of CM off-diagonal blocks in scalar field vacuum~\cite{gao2024detectingspacelikevacuumentanglement}, the separable identification of $\xi = 0$ can be further simplified.  In this case, the $X$ of separable CMs will reside in the $DCV$ intersection directly defined by $DCV_{\mp}$, i.e., the doubly-enclosed $DCV$ with upper(lower) apexes at $DCV_-^{(u)}$ and $DCV_{+}^{(\ell)}$ on the left(right) of Eq.~\eqref{eq:twodcv} (e.g., illustrated in Appendix Fig.~\ref{fig:coneconediagram2}).  

\subsection{Extremal spacelike field entanglement}
\label{sec:IIIB}

\begin{figure*}
 \begin{minipage}{0.9\textwidth}
 \includegraphics[width = \textwidth]{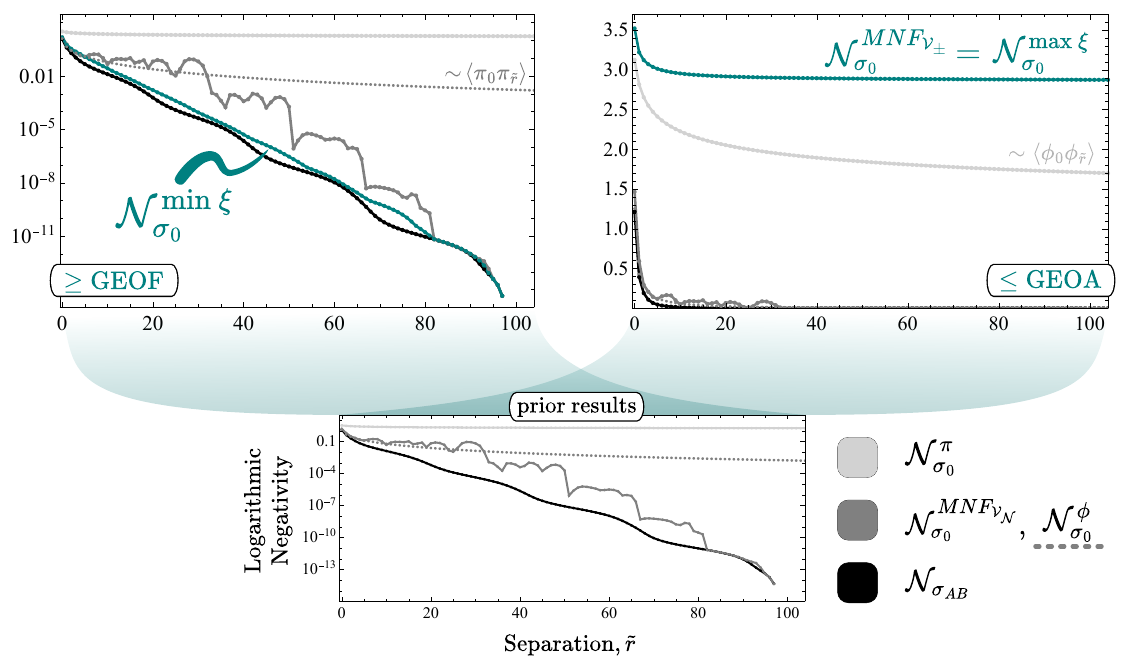}
  \end{minipage}\\
  \caption{Logarithmic negativity as a function of the separation $\tilde{r}$ between $AB$ in an infinite one-dimensional massless free lattice scalar field vacuum $(m, d) = \left( 10^{-10}, 10\right)$. Previous results of the logarithmic negativity upon volume tracing~\cite{scalar1dextra,NKnegativitysphere,NKentsphere} ($\mathcal{N}_{\sigma_{AB}}$, black), MNF-identified pure state~\cite{gao2024partialtransposeguided} and pure state identified through field-basis measurement~\cite{NKvolumemeasure} ($\mathcal{N}_{\sigma_{0}}^{\text{MNF}_{\mathcal{V}_{\mathcal{N}}}}$ and $\mathcal{N}_{\sigma_{0}}^{\phi}$, dark gray), and pure state identified through momentum-basis measurement~\cite{NKvolumemeasure} ($\mathcal{N}_{\sigma_{0}}^{\pi}$, light gray) are shown in all three panels for context. At this resolution, the field regions are separable at and beyond a dimensionless separation $\tilde{r}_{\Nslash} = 98$. The light(dark) gray indicate previous lower(upper) bounds to the GEOA(GEOF), each of which are surpassed by results of this work presented in teal. (Left) Tight upper bounds on GEOF. In a log-scale plot, highlighting the exponential decay with increasing separation, $\mathcal{N}_{\sigma_{0}}^{\min \xi}$ depicts the logarithmic negativity of the pure state identified through minimizing distance $\xi$ between $DCV_+$ and $DCV_-$. (Right) Tight lower bounds on GEOA. In a linear-scale plot, highlighting the double-logarithmic decay with increasing separation, $\mathcal{N}_{\sigma_{0}}^{\text{MNF}_{\mathcal{V}_\pm}} = \mathcal{N}_{\sigma_{0}}^{\max \xi}$ depicts the logarithmic negativity of the pure state identified through maximizing $\xi$, or alternatively through the generalized MNF procedure (Appendix~\ref{sec:appcsec3}).}
  \label{fig:plot}
\end{figure*}

Toward GEOA calculation, maximizing distance $\xi$ between $DCV_+$ and $DCV_-$ corresponds to $X_A \pm X_{AB} > 0$ residing at the upper(lower) apex of $DCV_{\pm}$ (see Appendix~\ref{sec:appcsec3} for further detail). To see the long distance entanglement behavior of the pure state given by this maximized solution (Eq.~\eqref{eq:MNFpurestate}), consider the simplest case of one site in each field region ($d$=1) separated by distance $\tilde{r}$ in lattice units. Since $X_{AB}\geq 0$, this two-mode logarithmic negativity is,
\begin{equation}
\mathcal{N}^{\text{max } \xi}_{\sigma_0, (1\times1)} =  -\log_2  \sqrt{\frac{(\sigma_{\pi}^{-1})_{A}  - (\sigma_{\pi}^{-1})_{AB}}{\sigma_{\phi,A} + \sigma_{\phi,AB}}} \ \ \ .
\label{eq:underlyingnegativity}
\end{equation}
The behavior of this expression at large separations differs between massive and massless fields. In the massive case, an asymptotic analysis of the two-point functions at large $\tilde{r}$~\cite{scalarvacuumoriginal1,NKvolumemeasure} reveals the underlying entanglement to follow $\mathcal{N}^{\text{max } \xi}_{\sigma_0, (1\times1)} \to \frac{1}{2\ln \left[2\right]}\left(- \ln \left[\frac{(\sigma_{\pi}^{-1})_{A}}{\sigma_{\phi,A}}\right]+\frac{(\sigma_{\pi}^{-1})_{AB}}{(\sigma_{\pi}^{-1})_{A}} + \frac{\sigma_{\phi,AB}}{\sigma_{\phi,A}}\right)$, i.e.,  decaying first exponentially before converging to a nonzero fixed value. For massless fields where asymptotic separations must remain within the Compton wavelength of the regulator in the regime $m\tilde{r}\ll 1$, an asymptotic analysis at large $\tilde{r}$ reveals the underlying entanglement to follow $\mathcal{N}^{\text{max } \xi}_{\sigma_0, (1\times1)} \to \frac{1}{\ln \left[4\right]}\Bigl( \ln \left[\left( \frac{4}{1-4\tilde{r}^2} - \pi \sigma_{\pi,A} \right)\left(\gamma - \pi \sigma_{\phi,A}+\ln \left[\frac{m\tilde{r}}{2}\right] \right)\right] - $ $  2 \ln \left[\pi\right] \Bigl)$, where $\gamma$ is the Euler-Mascheroni constant, i.e., decaying first polynomially before transitioning to a double-logarithmic decay at large distances. In the right panel of Fig.~\ref{fig:plot}, the corresponding behavior is shown in teal for a system with larger $d$, which is further along the trajectory towards the continuum.

Applying Weyl's inequality~\cite{hornmatrixana} to PT symplectic eigenvalues~\cite{Giedkepurestatetrans,gao2024partialtransposeguided} in multimode regimes (generalizing the two-mode discussion of Appendix~\ref{sec:appbsec1}), the maximization of underlying pure-state entanglement can be upper bounded by the sum of inverses of the $d$ largest PT symplectic eigenvalues. This solution would isolate the mixed-state's noise to the complementary space,  corresponding to the $d$ smallest PT symplectic eigenvalues. In order for this solution to be available, the former subspace (largest eigenvalues) must exhibit symplectic orthogonality among $A,B$-local components of its PT eigenvectors (SOL). Originally observed in the negativity-contributing subspace allowing consolidation of multimode entanglement into $(1_A \times 1_B)$ pairs~\cite{NKcorehalo,gao2024partialtransposeguided},  this requirement is a generalization of the SOL property to arbitrary subspaces of the PT eigensystem. The result that subspace SOL is necessary and sufficient to identify pure Gaussian state decompositions with Gaussian noise isolated to the SOL-complementary space is further discussed in Appendix~\ref{sec:appbsec2}. This upper bound is saturated by symmetric two-mode Gaussian states. However, the two-region reduced CM of the scalar field vacuum $\sigma_{AB}$ is found to violate SOL in the subspace of its largest PT symplectic eigenvalues. As such, the maximum GEOA between spacelike field regions is strictly excluded from this upper bound. For the right panel of Fig.~\ref{fig:plot}, this unreachable upper bound is approximately three times higher ($\sim 8.3$) than the reported lower bound at asymptotic separations.

Though the SOL subspace of the GEOA-saturating pure state cannot be made to coincide with the entire set of largest PT symplectic eigenvalues, the $\xi$-maximized pure state $\sigma_{0}^{\text{max } \xi}$ does include the largest one. Physically, this translates to $\sigma_{0}^{\text{max } \xi}$ containing the $(1_A \times 1_B)$ mode pair with the largest possible two-mode purified negativity. The sequential measurement protocol developed in Section~\ref{sec:IIA} then provides a derivation of two projective measurements in $C$ that would selectively purify this maximal two-mode entanglement resource (see first pair of measurement profiles in Appendix Fig.~\ref{fig:fig4} for an explicit example). Furthermore, the full SOL subspace of $\sigma_{0}^{\text{max } \xi}$ is found to be exactly $\mathcal{V}_{\pm}$, familiar from the derivation in Ref.~\cite{gao2024detectingspacelikevacuumentanglement} of the negativity's necessary and sufficient role in quantifying scalar field vacuum entanglement (see Appendix~\ref{sec:appbsec3} and~\ref{sec:appcsec3} for further detail). Because $\mathcal{V}_{\pm} \subset \mathcal{V}_{\Nslash}$ is contained in the negativity non-contributing subspace, every pair of communicated $C$ measurements organized by the $\mathcal{V}_{\pm}$ rank-one decomposition purifies an entangled ($\mathcal{N} > 0$) core pair in $AB$. This sequence of measurements provides an exponential hierarchy on the purified entanglement of $(1_A \times 1_B)$ pairs in $\sigma_{0}^{\text{max } \xi}$, as familiar from the behavior of the logarithmic negativity components~\cite{NKcorehalo,gao2024partialtransposeguided,gao2024detectingspacelikevacuumentanglement} and demonstrated with explicit projective measurement profiles in Appendix~\ref{sec:appbsec3} for a finite-volume scalar field vacuum. 

Since distance defined in a vector space is a convex function, $\xi$ can alternatively be minimized using standard convex optimization techniques, such as semidefinite programming (SDP). By numerically evaluating the $\xi$-minimization (details in Appendix~\ref{sec:appcsec4}) using the arbitrary-precision, parallelized semidefinite program solver SDPB~\cite{simmons2015semidefinite,Landry:2019qug}, the exponential decay of underlying pure-state entanglement with respect to separation between two disjoint regions is officially revealed. This exponential decay for massless free scalar field vacuum is shown by the teal line in the left panel of Fig.~\ref{fig:plot}, demonstrating a significant reduction in the upper bound of GEOF compared with previous estimates~\cite{NKvolumemeasure,gao2024partialtransposeguided}. For massive fields, the GEOF upper bound is observed to decay as a quadratic exponential, again following the logarithmic negativity. These results indicate that the amount of entanglement required to prepare a mixed state indistiguishable from the one observed by spacelike separated field detectors can be exponentially smaller than that present between the field regions, e.g., through communication of spatially local volume measurements~\cite{NKvolumemeasure}.

\section{Summary and Outlook}
\vspace{0.2cm}

This work establishes a unified Gaussian assisted paradigm for GEOA and GEOF through a direct correspondence between noise decompositions of mixed Gaussian states and hierarchies of projective measurements in an external purification. Framing calculations in the former language, a pair of double-cone volumes ($DCV$s) are introduced that provide a multimode geometric characterization of viable pure states underlying the noise. For disjoint regions of massless(massive) free scalar field vacuum, a distance metric $\xi$ between this $DCV$ pair is shown to be a necessary and sufficient entanglement quantifier. Governing optimizations by $\xi$ enables the discovery of strongest known bounds for entanglement measures with operational implications: a double-logarithmic(asymptotically constant) lower bound of GEOA and a linear(quadratic) exponentially decaying upper bound of GEOF with increasing spacelike separation. These bounds are seen to closely approximate their optimal values. Beyond scalar fields, the developed techniques serve to further enhance the outstanding computability of Gaussian quantum information.

Preliminary numerical results beyond those presented above indicate that the lower bound to GEOA, even at large separation ($\tilde{r}$), converges to a non-zero value as the resolution of the field ($d$) increases. This is consistent with observations that field regions are non-separable at all distances in the continuum~\cite{{scalar1dextra,NKnegativitysphere,NKentsphere,gao2024detectingspacelikevacuumentanglement}}, i.e., $GEOA \geq GEOF > 0$. Thus, there may exist a finite entanglement resource in the massive continuum limit at infinite separation through the assisted protocol. The analytically provided $\xi$ maximization will allow extrapolation of this value,  and the emergence of its distance-invariance, into the continuum.

Given an $ABC$ pure state, the constructive procedures of this manuscript provide an exact calculation of $C$ measurements capable of removing noise $Y = \sigma_{AB} - \sigma_0$ from $AB$. While the language of ideal quantum operations and arbitrary-precision numerical environments have allowed theoretical clarity, continuing the current trajectory toward experimental settings will require generalizing techniques to incorporate realistic resources, e.g., noise in the $ABC$ global state, only partial access to the $C$ purification, or measurement imperfections. In such scenarios, the observed sensitivity of $AB$-purified entanglement to the $C$-measurement profile may also necessitate great care in experimental design.

The inter-$DCV$ distance metric introduced in this work is found to alleviate difficulties in calculating multimode entanglement properties of mixed Gaussian states, which commonly involve non-convex optimizations over all viable pure-state decompositions. However, addressing entanglement measures associated with constrained optimizations, which cannot directly utilize globally optimal strategies developed here in the assisted setting, remains a target of future innovations. One closely related example is the Localizable Entanglement (LE)~\cite{Verstraete:2003spp,localizableentanglement}, which quantifies the maximum $AB$ quantum resources that can be identified through only local measurements in the $C$ purification ($U = \bigotimes_{j = 1}^{N_C} U_j$ in Fig.~\ref{fig:diagram}) and provides a route for defining an entanglement length. Whether the $\phi$- and $\pi$-basis measurements~\cite{NKvolumemeasure} are extremal in this context of local measurements is currently unknown. The present $DCV$ framework demonstrates the existence, and inspires further creation, of strategic quantifiers for pursuing many-body entanglement structure beyond guidance of the PT eigensystem.

Extending previous two-mode observations~\cite{PhysRevA.101.052305} in completely uncorrelated and GLEMS~\cite{adesso2004determination,It1} Gaussian states, those with AB symmetry are now also shown to have optimal GEOA solutions that geometrically reside on the two-mode $DCV$ rim. Additional numerical studies performed for arbitrary two-mode Gaussian states support a conjecture that the optimal solution is generally a member of this one-dimensional subspace. If analytically confirmed, this would deepen the duality presently realized between GEOA and GEOF optimizations.

The techniques of $\xi$-minimization toward GEOF have determined that the pure-state entanglement required to produce the mixed density matrix of spacelike field regions decays exponentially with separation, following the logarithmic negativity. This is substantially smaller than the entanglement present between these regions, e.g., as uncovered by classical communication from the external volume with LE lower bounded by correlation functions~\cite{Verstraete:2003spp,localizableentanglement}. In the spirit of strategically incorporating quantum noise in experimental protocols~\cite{Terhal:1998ci,RevModPhys.93.015008,Ezzell:2022net,Than:2024zaj}, this result reveals potential for reducing quantum resource requirements for simulating quantum fields~\cite{Jordan:2012xnu,Jordan:2011ci,Jordan:2017lea} by designing simulations from the perspective of local detectors, as observed by collider experiments, rather than the perspective of the omniscient full volume.

\begin{acknowledgments}
We thank D.~H.~Beck, Edward Chen, Iman Marvian, Enrique Rico, and Michael M.~Wolf for insightful discussions during the development of this work. We thank also participants of the \emph{Entanglement in Many-Body Systems: From Nuclei to Quantum Computers and Back} workshop and \emph{Co-design for Fundamental Physics in the Fault-Tolerant Era} workshop at the InQubator for Quantum Simulation (IQuS) hosted by the Institute for Nuclear Theory (INT) for interactions. IQuS is supported by U.S. Department of Energy, Office of Science, Office of Nuclear Physics, under Award Number DOE (NP) Award DE-SC0020970 via the program on Quantum Horizons: QIS Research and Innovation for Nuclear Science, and by the Department of Physics, and the College of Arts and Sciences at the University of Washington. BG is supported in part by the Goshaw Family Endowment fellowship and Trinity Summer Graduate Assistantship. Numerical computations were carried out using the Mathematica 14 arbitrary precision libraries~\cite{mma} with up to 5000 digits of precision, Duke Computing Cluster (DCC), and semidefinite programming solver SDPB~\cite{simmons2015semidefinite,Landry:2019qug}.
\end{acknowledgments}

\section*{Data Availability}
The numerical values that support the findings of this article are openly available~\cite{zenodovalues}.

\bibliography{biblio}

\onecolumngrid
\newpage
\appendix

\section{Projective measurement description of Gaussian noise decompositions}
\label{sec:appa}

The key result of this appendix is a constructive procedure for connecting any valid Gaussian noise decomposition of a reduced CM, i.e., $\text{Tr}_C \left[ \rho_\sigma \right] \equiv \rho_{\sigma_{AB}}$, with $\sigma_{AB} = \sigma_0 + Y$, $Y\geq 0$, and $\sigma_0 +i \Omega \geq 0$, to a projective measurement protocol in the $C$ Hilbert space with available auxiliary modes. Regarding $Y$ as a sum of rank-one components, $\{ Y_1\}$, Appendices~\ref{sec:appasec3} and~\ref{sec:appasec4} identify a collective operator profile $\langle p|$ in $C$ (in the present application, $C$ is the field volume outside detection regions $A$ and $B$) whose projective measurement (i.e., process of Fig.~\ref{fig:diagram}) removes a $Y_1$ from the reduced CM,
\begin{equation}
    Y_{1} = \sigma_{AB} - \left( \left( \left(  \mathbb{I}_{AB} \oplus S_{\langle p|} \right)    \sigma \left( \mathbb{I}_{AB} \oplus S_{\langle p|} \right)^{T} \right)^{'} \right)_{AB} \equiv \sigma_{AB} - \sigma^{\quad  '}_{p,AB} \ \ \ ,
    \label{eq:labcons}
\end{equation}
where $S_{\langle p|}$ acts in the $C$ space to place the linear combination of phase space operators governed by the collective profile onto the $\phi$-component of one mode, which is then projectively measured in the $\phi$-basis as denoted by the prime. Placing the normal mode profile $\langle p |$ onto one mode, the state prior to measurement becomes $\sigma_p \equiv \left(  \mathbb{I}_{AB} \oplus S_{\langle p|} \right)    \sigma \left( \mathbb{I}_{AB} \oplus S_{\langle p|} \right)^{T}$. Along with classical communication of the measurement result, this describes procedures in the lab capable of removing $Y_1$ from the $AB$ reduced CM.

In order to find $S_{\langle p|}$ for any given $Y_{1}$, it is advantageous to rewrite Eq.~\eqref{eq:labcons} in the basis of local normal modes for the bipartition $AB|C$, where the global pure state $\sigma$ simplifies to a tensor product of two-mode squeezed vacuum states (TMSVSs)~\cite{PhysRevA.67.052311, Giedkepurestatetrans}, achieved through local Williamson transformation $\bar{\sigma} = \left(\bar{S}_{AB} \oplus \bar{S}_{C}\right) \sigma \left(\bar{S}_{AB} \oplus \bar{S}_{C}\right)^T$. The noise in the original basis connects to that in the basis of local normal modes $\bar{Y}$ via,
\begin{align}
Y_{1}  &= \sigma_{AB} - \left( \left( \left( \bar{S}_{AB}^{-1} \oplus S_{\langle p|}\bar{S}_{C}^{-1} \right)   \bar{\sigma} \left( \bar{S}_{AB}^{-1}\oplus S_{\langle p|}\bar{S}_{C}^{-1} \right)^{T} \right)^{'} \right)_{AB} \label{eq:commuteMandT0} \\ &= \bar{S}_{AB}^{-1} \left(  \bar{\sigma}_{AB}  - \left( \left( \left( \mathbb{I}_{AB} \oplus S_{\langle \bar{p} |} \right)   \bar{\sigma} \left( \mathbb{I}_{AB} \oplus S_{\langle \bar{p} |} \right)^{T} \right)^{'} \right)_{AB} \right)\bar{S}_{AB}^{-T} \label{eq:commuteMandT} \\ &= \bar{S}_{AB}^{-1} \bar{Y}_{1} \bar{S}_{AB}^{-T} \label{eq:commuteMandT1} \ \ \ ,
\end{align}
where the measurement in $C$ commutes with any local symplectic (local unitary in the Hilbert space) transformations on AB in Eq.~\eqref{eq:commuteMandT}. Procedures for calculating $\bar{S}_{AB}$ and $\bar{S}_{C}$ that perform the local symplectic diagonalization are discussed, for example, in Appendix D of Ref.~\cite{NKvolumemeasure} or Appendix B2 of Ref.~\cite{gao2024partialtransposeguided}. As such, Eqs.~\eqref{eq:commuteMandT0}--\eqref{eq:commuteMandT1} reformulate the problem as finding $S_{\langle \bar{p} |}$ for any given $\bar{Y}_{1}$. Because $\bar{\sigma}$ is simply a set of TMSVSs,  it is convenient to first determine $S_{\langle \bar{p} |} = S_{\langle p|} \bar{S}_{C}^{-1}$ for a given $\bar{Y}$ as an intermediate step to the desired $S_{\langle p|}$ associated with a given $Y_1$.

The translation of Eqs.~\eqref{eq:commuteMandT0}--\eqref{eq:commuteMandT1} takes a general global pure state $\sigma$ to its associated $\phi\pi$-uncorrelated local normal modes $\bar{\sigma}$. Appendix~\ref{sec:appasec1} will first focus on measurement profiles that are also $\phi\pi$-uncorrelated in this basis, building an analytic formula connecting the measurement of a $\phi\pi$-uncorrelated collective operator profile $\langle\bar{p}^\phi|$ to the resulting noise subtracted from the reduced CM. In Appendix~\ref{sec:appasec2} this profile is generalized to be $\phi\pi$-correlated, $\langle \bar{p} |$, extending the established analytic connection to the most general scheme of projective measurement. With the analytics established in the basis of local normal modes, the desired reverse process is developed in Appendices~\ref{sec:appasec3} and~\ref{sec:appasec4}. Appendix~\ref{sec:appasec3} determines a measurement procedure given any  noise, incorporating auxiliary modes in $C$ when necessary. This enables a procedure that constructs a sequence of projective measurements in $C$ for any given  decomposition of noise in Appendix~\ref{sec:appasec4}. In the lab, this can be implemented as a single symplectic transformation in $C$ followed by projectively measuring the last $\rank \left( Y \right)$ modes of C in the $\phi$ basis, depicted in Fig.~\ref{fig:circuitdiagram}. As discussed in Sec.~\ref{sec:IIA}, the procedures discussed in this appendix clarify the operational meaning of Gaussian classical mixing channels, $\sigma_{AB} = \sigma_0 + Y$, for which the entanglement in $\sigma_0$ is extremized in the calculation of the GEOF and GEOA.

\subsection{Rank-one noise removal given a \texorpdfstring{$\phi\pi$}{phipi}-uncorrelated projective measurement}
\label{sec:appasec1}

This section performs a measurement on $C_2$ that is $\phi\pi$-uncorrelated in the local normal mode basis. An analytic expression is derived relating the $C$ collective measurement profile and the resulting noise removed in $AB$.

As discussed in Eqs.~\eqref{eq:commuteMandT0}--\eqref{eq:commuteMandT1}, when analyzing measurement effects for a particular CM, it is advantageous to work from a perspective of local normal modes, which may be parametrized as,
\begin{equation}
  \scalemath{0.96}{\bar{\sigma} =   \begin{pmatrix}
        D_{AB, cosh} & D_{ABC, sinh} & 0 & 0 \\
        D_{ABC, sinh}^{T} & D_{C, cosh} & 0 & 0 \\
        0 & 0 & D_{AB, cosh}  & -D_{ABC, sinh}  \\
        0 & 0 & -D_{ABC, sinh}^{T} & D_{C, cosh} \\
    \end{pmatrix}, \quad (\bar{\sigma}_{AB}^{})^{-1} = \begin{pmatrix}
        D_{AB, cosh} & 0 \\
        0 & D_{AB, cosh} \\
    \end{pmatrix}^{-1} 
    \label{eq:normalform1}}  ,
\end{equation}
where $D_{X, cosh}$ and $D_{X, sinh}$ are matrices governed by two-mode squeezing parameters (and thus local symplectic eigenvalues) in the form of $\cosh (2 r_j)$ and $\sinh (2 r_j)$ on the diagonal, with $j$ ranging from 1 to the number of modes in each subspace. Following the main text, all $\phi$ components are placed before the corresponding $\pi$ components in the canonical operator ordering. When the number of modes in $AB$, $N_{AB}$, is different from that in $C$, $N_C$, the party with a larger number of modes will have pure single normal modes decoupled from the rest of the system, resulting in an $N_{AB} \times N_{C}$ rectangular matrix for $D_{ABC,sinh}$. Without loss of generality, positively squeezed TMSVS pairs ($r_j \geq 0$) are chosen~\footnote{Any negatively squeezed TMSVS can be transformed to its positively squeezed counterpart via application of a single-mode phase shifter with angle $\pi$.} in designing the normal mode symplectic transformation resulting in Eq.~\eqref{eq:normalform1}. With this transformation, any CM can be written in a $\phi\pi$-uncorrelated form.

To determine the noise removed from $\bar{\sigma}$ through a $\phi\pi$-uncorrelated projective measurement, consider a $\phi$-basis measurement on a general $\phi\pi$-uncorrelated CM, $\sigma^{\phipislash}$. In terms of the two-point correlators of the conjugate momenta, $K_{ij} = 2 \langle \hat{\pi}_i \hat{\pi}_j \rangle$, an arbitrary $\phi\pi$-uncorrelated pure state CM may be composed as 
\begin{equation}
\sigma^{\phipislash} = X \oplus X^{-1} \quad , \quad  X^{-1} \equiv \begin{pmatrix}
K_{AB} & K_{ABC_{1}} & K_{ABC_{2}} \\
K_{ABC_{1}}^{T} & K_{C_{1}} &  K_{C_{1}C_{2}} \\
K_{ABC_{2}}^{T} & K_{C_{1}C_{2}}^{T} &  K_{C_{2}} \\
\end{pmatrix} \ \ \ .
\end{equation} 
In the present application, $AB$ is the subspace where Gaussian entanglement resources will be concentrated and the union $A\cup B\cup C_1 \cup C_2$ is a pure state. When $\sigma^{\phipislash}$ is partially projected onto a particular field-space configuration in $C_2$, the CM of the resulting pure states is governed by the principal submatrix of $X^{-1}$ corresponding to the remaining modes~\cite{NKvolumemeasure}. The mixed $AB$ CM after $\phi$ measurement in $C_2$ reads,
\begin{equation}
\sigma_{\quad AB}^{\phipislash \quad '}  = \begin{pmatrix}
\left( \left( (X^{-1})_{ABC_{1}} \right)^{-1} \right)_{AB} & 0 \\
0 & (X^{-1})_{AB} \\
\end{pmatrix} \ \ \ , 
\label{eq:paramofXinv}
\end{equation}
where different measurement outcomes result in the same CM and different first-order displacements, which can be unified through classical communication of the measurement result~\cite{NKvolumemeasure}. The difference of CM inverses before and after projective Gaussian measurement on $C_{2}$ reads,
\begin{equation}
    (\sigma_{\quad AB}^{\phipislash \quad '})^{-1} - (\sigma_{AB}^{\phipislash})^{-1} =  \begin{pmatrix}
\begin{pmatrix}
K_{ABC_{1}}  & K_{ABC_{2}} \\
\end{pmatrix} \left( \begin{pmatrix}
K_{C_{1}} &  K_{C_{1}C_{2}} \\
 K_{C_{1}C_{2}}^{T} &  K_{C_{2}} \\
\end{pmatrix}^{-1} -  \begin{pmatrix}
K_{C_{1}}^{-1} &  0 \\
 0 &  0 \\
\end{pmatrix} \right ) \begin{pmatrix}
K_{ABC_{1}}^{T}  \\
K_{ABC_{2}}^{T} \\
\end{pmatrix}   & 0 \\
0 & 0 \\
\end{pmatrix}  \ \ \ .
\label{eq:diffinv}
\end{equation}
When $C_2$ corresponds to a single mode, the inner matrix is $K_{C}^{-1} - \left(K_{C_1}^{-1} \oplus 0\right) \equiv v |v\rangle \langle v|$, where $|v\rangle = \frac{1}{v} \begin{pmatrix}
-K_{C_{1}}^{-1}K_{C_{1}C_{2}} & 1  \\
\end{pmatrix}^{T}$ is the (unnormalized) last row/column of $K_{C}^{-1}$ and $v =  K_{C_2} - K_{C_{1}C_{2}}^{T} K_{C_1}^{-1} K_{C_{1}C_{2}} \geq 0$ is the (scalar) Schur complement.

From the normal-mode perspective, a $\phi$-basis collective operator with linear combination governed by the measurement profile, $\langle \bar{p}^\phi|$, can be placed on the $\phi$ phase space operator of $C_2$ via a $\phi\pi$-uncorrelated symplectic transformation in $C$, $S_{\langle \bar{p}^\phi|}  \equiv  \begin{pmatrix} L & 0 \\ 0 & L^{-T} \end{pmatrix}$, constructed such that $\langle \bar{p}^{\phi}|$ is the last row of $L$. In this context, Eq.~\eqref{eq:diffinv} applies through the relations,
\begin{equation}
    \begin{pmatrix}
K_{C_{1}} &  K_{C_{1}C_{2}} \\
 K_{C_{1}C_{2}}^{T} &  K_{C_{2}} \\
\end{pmatrix}  = L^{-T} D_{C,cosh} L^{-1}, \quad  \begin{pmatrix}
K_{ABC_{1}}  & K_{ABC_{2}} \\ 
\end{pmatrix} = -D_{ABC,sinh}L^{-1} \ \ \ .
\label{eq:design1}
\end{equation}
With $D_{ABC,tanh}\equiv D_{ABC,sinh}D_{C,cosh}^{-1}$,
\begin{equation}
\begin{pmatrix}
K_{ABC_{1}}  & K_{ABC_{2}} \\ 
\end{pmatrix} | v \rangle = - D_{ABC,tanh} | \bar{p}^{\phi} \rangle ,\quad v =  \frac{1}{\langle \bar{p}^{\phi} | D_{C,cosh}^{-1} | \bar{p}^{\phi} \rangle} \ \ \ .
\label{eq:diffinv1}
\end{equation}
Because the profile symplectic acts only in the $C$ space, $\bar{\sigma}_{\bar{p}^{\phi},AB} = \bar{\sigma}_{AB}$. As such,
\begin{equation}(\bar{\sigma}_{\bar{p}^{\phi},AB}^{ \quad '})^{-1} - (\bar{\sigma}_{\bar{p}^{\phi},AB}^{})^{-1} =  \begin{pmatrix}
\frac{1}{\langle \bar{p}^\phi | D_{C,cosh}^{-1} | \bar{p}^\phi \rangle}  D_{ABC,tanh} | \bar{p}^\phi  \rangle \langle \bar{p}^\phi | D_{ABC,tanh}^{T}  & 0 \\
0 & 0 \\
\end{pmatrix} \equiv \bar{v}_{\phi} | \bar{v}_{\phi} \rangle \langle \bar{v}_{\phi} | \ \ \ ,
\label{eq:diffinv2}
\end{equation}
with the eigenvector and eigenvalue,
\begin{equation}
    | \bar{v}_{\phi} \rangle = \frac{1}{\sqrt{\langle \bar{p}^\phi | D_{ABC,tanh}^{T} D_{ABC,tanh}| \bar{p}^\phi \rangle}} \begin{pmatrix}
 D_{ABC,tanh}  | \bar{p}^\phi \rangle \\
0 \\
\end{pmatrix}, \quad  \bar{v}_{\phi} = \frac{  \langle \bar{p}^\phi | D_{ABC,tanh}^{T} D_{ABC,tanh}| \bar{p}^\phi \rangle }{\langle \bar{p}^\phi | D_{C,cosh}^{-1} | \bar{p}^\phi \rangle}\ \ \ .
\label{eq:re1}
\end{equation}
Given any $ABC$ pure state and $\phi\pi$-uncorrelated measurement profile among the $C$ local normal modes, Equations~\eqref{eq:diffinv2} and~\eqref{eq:re1} can be solved for the resulting CM (left of Eq.~\eqref{eq:diffinv2}), and thus for the consequently removed noise. Note that this expression is independent of the overall normalization of $\langle\bar{p}^\phi|$, which is a freedom that can be understood as insensitivity to a single-mode squeezing operation applied to $C_2$ prior to measurement. Similar analysis carries over when $\bar{\sigma}$ is partially projected onto a particular $\pi$-space configuration in the volume. 

\subsection{Rank-one noise removal given a \texorpdfstring{$\phi\pi$}{phipi}-correlated projective measurement}
\label{sec:appasec2}

This section generalizes the establishment of Appendix~\ref{sec:appasec1} to projective measurement with $\phi\pi$-correlated collective operators in $\bar{\sigma}$. Because measurement is performed on the single site where the collective operator has been localized, the general symplectic transformation in $C$ that performs this localization may be expressed simply by introducing an initial layer of single-mode phase shifts,
\begin{equation}
     R_{C} \equiv \begin{pmatrix}
       \diag \left( \cos{\theta_{j}} \right) & \diag \left( \sin{\theta_{j}} \right) \\
      \diag \left( -\sin{\theta_{j}} \right) & \diag \left( \cos{\theta_{j}} \right) \\
    \end{pmatrix}, \quad \langle \bar{p}^{\phi} | \equiv \begin{pmatrix}
        \langle \bar{p}^{\phi} | & 0 \\
    \end{pmatrix}, \quad \langle \bar{p} | \equiv \langle \bar{p}^{\phi} | R_{C},\quad S_{\langle \bar{p} |} \equiv  S_{\langle \bar{p}^{\phi} |} R_{C} \ \ \ ,
    \label{eq:fromofR}
\end{equation}
where again $j$ indexes the modes in $C$, and the corresponding row vector of $S_{\langle \bar{p} |}$ multiplied by $R_{C}$ provides sufficient degrees of freedom to construct a general $\phi\pi$-correlated collective profile $ \langle \bar{p} |$. Constructing $S_{\langle \bar{p}|}$ from $\langle \bar{p}|$ may be achieved through a symplectic Gram-Schmidt (sGS) procedure, beginning with $\langle\bar{p}|$ as the row governing the linear combination placed on the $\phi$ space of the last mode in $C$. When projectively measuring this last mode in the $\phi$-basis, the measurement results in $AB$ reflect only the profile vector, independent of the residual degrees of freedom in the sGS procedure. Capturing this freedom through the sGS procedure simplifies the analytic connection between a $\phi\pi$-correlated collective measurement profile and the resulting noise reduction.

Due to the symmetry of single-mode phase shifts on TMSVS states, the $\phi\pi$-correlation introduced by application of $S_{\langle \bar{p}|}$ to the $C$-space of $\bar{\sigma}$ can be equivalently achieved by applying $S_{\langle \bar{p}|} = R_{AB}^{T} \oplus S_{\langle \bar{p}^\phi|}$, where the phase shifts have been relocated~\footnote{If $N_{C} \geq N_{AB}$, $R_{AB} = (R_{C}^{T})_{AB}$. If $N_{C} < N_{AB}$, $R_{AB} = R_C^T \oplus \mathbb{I}$, where the identity has dimension corresponding to the subspace of pure normal modes of AB that decouple from the rest of the system.} to the $AB$ space. Through commuting operations as in Eq.~\eqref{eq:commuteMandT}, the symplectic transformation in $AB$ can be delayed until after the measurement occurs in the $\phi\pi$-uncorrelated picture,
\begin{equation}    (\bar{\sigma}_{\bar{p},AB}^{\quad '})^{-1} - (\bar{\sigma}_{\bar{p},AB}^{})^{-1} = R_{AB}^{T}  \left( (\bar{\sigma}_{\bar{p}^{\phi},AB}^{\quad '})^{-1} - (\bar{\sigma}_{\bar{p}^{\phi},AB}^{})^{-1}\right) R_{AB} \ \ \ .
    \label{eq:askforsimpli}
\end{equation}
From Eq.~\eqref{eq:diffinv2}, with $R_{AB}^{T} \left( D_{ABC,tanh} \oplus -D_{ABC,tanh} \right)  | \bar{p}^{\phi} \rangle =  \left( D_{ABC,tanh} \oplus -D_{ABC,tanh} \right) | \bar{p} \rangle$ and $\langle \bar{p}^{\phi} | \left(D_{C,cosh}^{-1} \oplus D_{C,cosh}^{-1} \right) | \bar{p}^{\phi} \rangle =\langle \bar{p} |\left(D_{C,cosh}^{-1} \oplus D_{C,cosh}^{-1} \right) | \bar{p} \rangle$,
\begin{equation}
    \scalemath{0.96}{(\bar{\sigma}_{\bar{p},AB}^{\quad '})^{-1} - (\bar{\sigma}_{\bar{p},AB}^{})^{-1}  =   \frac{ \left( D_{ABC,tanh} \oplus -D_{ABC,tanh} \right) | \bar{p} \rangle \langle \bar{p} | \left( D_{ABC,tanh} \oplus -D_{ABC,tanh} \right)^{T} }{\langle \bar{p} | \left(D_{C,cosh}^{-1} \oplus D_{C,cosh}^{-1} \right) | \bar{p} \rangle} \equiv \bar{v} | \bar{v} \rangle \langle  \bar{v} |}  \ \ \    ,
        \label{eq:conn0}  
\end{equation}
where eigenvector $| \bar{v} \rangle$ is,
\begin{equation}
    | \bar{v} \rangle = \frac{\left( D_{ABC,tanh} \oplus -D_{ABC,tanh} \right) | \bar{p} \rangle}{\sqrt{\langle \bar{p} | \left( D_{ABC,tanh} \oplus -D_{ABC,tanh} \right)^{T}  \left( D_{ABC,tanh} \oplus -D_{ABC,tanh} \right) | \bar{p} \rangle}} \ \ \ ,
    \label{eq:conn1}    
\end{equation}
and the corresponding eigenvalue $\bar{v}$ is,
\begin{equation}
 \bar{v} = \frac{\langle \bar{p} | \left( D_{ABC,tanh} \oplus -D_{ABC,tanh} \right)^{T}  \left( D_{ABC,tanh} \oplus -D_{ABC,tanh} \right) | \bar{p} \rangle}{\langle \bar{p} | \left(D_{C,cosh}^{-1} \oplus D_{C,cosh}^{-1} \right) | \bar{p} \rangle}\ \ \ .
    \label{eq:conn2}
\end{equation}
In the basis of local normal modes achieved through $\bar{S}_{AB} \oplus \bar{S}_{C}$, Eqs.~\eqref{eq:conn0}-\eqref{eq:conn2} provide the general relation from measurement profile to post-measurement reduced CM applicable 
for an arbitrarily correlated CM and profile.

In order to confirm this calculation performed in the basis of local normal modes with the corresponding process on the lattice when either $\sigma$ and/or $\langle \bar{p}|$ are $\phi\pi$-correlated, it is necessary to extend the formalism of Eq.~\eqref{eq:paramofXinv} for updating the reduced CM upon the present measurement and communication procedure. With $m$ the set of modes on which $\phi$-basis measurement is performed and $r$ the remaining system, an $m \cup r$ pure state of $N = [m]+[r]$ modes yields a projected state of the form,
\begin{equation}
|\psi_r\rangle_{\phi_m} \propto \langle \phi_m | \psi\rangle =  \frac{\det X^{-\frac{1}{4}}}{\pi^{\frac{N}{4}}} e^{-\frac{1}{2} \phi_m^T L_{mm} \phi_m} \\ \int \dif \phi_r \ e^{-\frac{1}{2} \phi_r^T L_{rr}  \phi_r - u^T\phi_r }|\phi_r\rangle \ \ \ ,
\end{equation}
where $L \equiv X^{-1} - i Y$ as parametrized in Eq.~\eqref{eq:vacuumWF}, $u = L_{rm} \phi_m$, and $\phi_{m}$ is the projected value of $\hat{\phi}_{m}$. With standard Gaussian integration, the normalized pure state reads,
\begin{equation}
|\psi_r\rangle_{\phi_m} =  \frac{\det  \left(\left(X^{-1}\right)_{rr}\right)^{\frac{1}{4}}}{\pi^{\frac{[r]}{4}}} e^{-\frac{1}{2} \text{Re}\left( u\right)^T  \left(\left(X^{-1}\right)_{rr}\right)^{-1} \text{Re}\left( u\right)} \\ \int \dif \phi_r \ e^{-\frac{1}{2} \phi_r^T L_{rr} \phi_r - u^T\phi_r }|\phi_r\rangle \ \ \ .
\label{eq:a41}
\end{equation}
Beginning from Eq.~\eqref{eq:pureCMform}, the CM for this pure state is independent of measurement outcome,
\begin{equation}
\sigma_{|\psi_r\rangle_{\phi_m}}  = \begin{pmatrix}
\left( \left( X^{-1}\right)_{rr} \right)^{-1} & \left( \left( X^{-1}\right)_{rr} \right)^{-1} Y_{rr} \\
Y_{rr} \left( \left( X^{-1}\right)_{rr} \right)^{-1} & Y_{rr} \left( \left( X^{-1}\right)_{rr} \right)^{-1} Y_{rr} + \left( X^{-1}\right)_{rr} \\
\end{pmatrix} \ \ \ ,
\label{eq:a42}
\end{equation}
and the measurement-dependent displacements of its first moments are,
\begin{subequations}
\begin{align}
&\langle \hat{\phi}_r \rangle = -\left( \left(X^{-1}\right)_{rr} \right)^{-1} \left(X^{-1}\right)_{rm}  \phi_m  \ \ \ ,  \\
&\langle \hat{\pi}_r \rangle = -  \left(    Y_{rm} -Z_{rr} \left(\left(X^{-1}\right)_{rr} \right)^{-1}Y_{rr} Z_{rr}^{-1} \left(X^{-1}\right)_{rm} \right)  \phi_m  \ \ \ ,
\end{align}
\label{eq:rfirstmoments}%
\end{subequations}
where $Z_{rr} \equiv  \left(X^{-1}\right)_{rr} + Y_{rr}\left(\left(X^{-1}\right)_{rr}\right)^{-1} Y_{rr} $. While all measurement outcomes in the $\phi$-basis of $m$ will result in the CM of Eq.~\eqref{eq:a42}, if the result of the measurement is unknown, uncertainty in the displacement will result in an $r$ mixed state in the particular convex decomposition,
\begin{equation}
  \rho_r =  \int  \dif \phi_m \ \langle \phi_m |\psi\rangle \langle \psi | \phi_m \rangle = \int \dif \phi_m \ |\mathcal{A}_{\phi_m}|^2 \ \rho\Big( |\psi_r\rangle_{\phi_m} \Big)\ \ \ ,
  \label{eq:a44}
\end{equation}
with the $\phi_m$-configuration-dependent amplitude $|\mathcal{A}_{\phi_m}|^{2} = \frac{1}{\pi^{\frac{[m]}{2}}} \frac{\det X^{-\frac{1}{2}}}{\det  \left(\left(X^{-1}\right)_{rr}\right)^{\frac{1}{2}}} \frac{e^{- \phi_m^T \left( X^{-1}\right)_{mm} \phi_m}}{e^{- \text{Re}\left( u\right)^T  \left(\left(X^{-1}\right)_{rr}\right)^{-1} \text{Re}\left( u\right)}}$.

\subsection{Constructing a projective measurement given a rank-one noise removal}
\label{sec:appasec3}

Providing the desired reversal of Appendix~\ref{sec:appasec1} and~\ref{sec:appasec2}, this section constructs $|\bar{p}\rangle$ governing a measurement protocol for any rank-one noise removal, systematically introducing auxiliary modes when necessary. Given any  noise decomposition, $\sigma_{AB}^{} = \sigma_{AB}^{'} + Y_{1}$, symplectic transformation to the basis of local normal modes by $\bar{S}_{AB}$ yields $\bar{\sigma}_{AB}^{} = \bar{\sigma}_{AB}^{'} + \bar{Y}_{1}$, where $\rank ( Y_{1} ) = \rank ( \bar{Y}_{1} ) = 1$, $\bar{\sigma}_{AB}^{}>0$ and $\bar{\sigma}_{AB}^{'}>0$. The inverse matrix subtraction is thus,
\begin{equation}
    (\bar{\sigma}_{AB}^{'})^{-1} - (\bar{\sigma}_{AB}^{})^{-1}  =\bar{v} \ket{\bar{v}}\bra{\bar{v}}= \frac{\left( \bar{\sigma}_{AB}\right)^{-1} \bar{Y}_{1} \left( \bar{\sigma}_{AB}\right)^{-1} }{1-\Tr \left[ \left( \bar{\sigma}_{AB}\right)^{-1} \bar{Y}_{1} \right]} \ \ \  ,
\label{eq:matrixmani} 
\end{equation}
utilizing Sherman–Morrison–Woodbury formula~\cite{10.1214/aoms/1177729893,10.1214/aoms/1177729698,doi:10.1137/1031049} and the cyclic property of the trace.

To determine the appropriate profile vector $| \bar{p} \rangle$ that ensures subtraction of $\bar{Y}_{1}$ through the measurement procedure, define,
\begin{equation}
   |\bar{p}_{0}\rangle \equiv \left( D_{ABC,tanh} \oplus -D_{ABC,tanh} \right)^{-1} |\bar{v}\rangle \ \ \ ,
       \label{eq:pveccons0}
\end{equation}
which by-construction satisfies Eqs.~\eqref{eq:conn1} and sets the numerator of Eqs.~\eqref{eq:conn2} to one. Note that $2 \left( N_{C} - N_{AB} \right)$ entries of $|\bar{p}_{0}\rangle$ are zero, which can be freely adjusted without affecting the equality in Eq.~\eqref{eq:conn1}. Since an overall scaling factor on $| \bar{p} \rangle$ leaves Eqs.~\eqref{eq:conn1} and~\eqref{eq:conn2} unchanged, the zero subspace of $| \bar{p}_{0} \rangle$ provides the only sensitivity to achieve the desired eigenvalue of Eq.~\eqref{eq:conn2} while maintaining Eq.~\eqref{eq:conn1}. When $N_C > N_{AB}$, this subspace may be composed of the set of pure optical-vacuum modes ($\sigma = \mathbb{I}$) in $C$ that decouple from the system upon transformation to the local normal modes; when $N_C \leq N_{AB}$, this subspace must be introduced as a pure auxiliary mode initialized to this vacuum state. For example, one way to construct a faithful profile vector $ | \bar{p} \rangle$ is,
\begin{equation}
   | \bar{p} \rangle = | \bar{p}_{0} \rangle + \begin{pmatrix}
       0 \\
      \sqrt{\frac{1}{\bar{v}} - \langle \bar{p}_{0} | \left(D_{C,cosh}^{-1} \oplus D_{C,cosh}^{-1} \right) | \bar{p}_{0} \rangle}  \\
     0 \\
   \end{pmatrix} \ \ \ ,
   \label{eq:pveccons}
\end{equation}
where on the right side of Eq.~\eqref{eq:pveccons}, the top zero vector has a length of $N_{AB}$, and the bottom zero vector has a length of $2 N_{C} - N_{AB}  -1$. To see that the argument of the square root is a non-negative valued function, note that,
\begin{align}
   \bar{v} \langle \bar{p}_{0} | \left(D_{C,cosh}^{-1} \oplus D_{C,cosh}^{-1} \right) | \bar{p}_{0} \rangle \label{eq:domainkey} &= \Tr \left[ \left( \bar{\sigma}_{AB}^{} - \left( \bar{\sigma}_{AB}\right)^{-1} \right)^{-1} \left( (\bar{\sigma}_{AB}^{'})^{-1} - (\bar{\sigma}_{AB}^{})^{-1}  \right) \right]   \\ &= \frac{\Tr \left[ \left( \bar{\sigma}_{AB}^{} - \left( \bar{\sigma}_{AB}\right)^{-1} \right)^{-1} \bar{Y}_{1} \right]-\Tr \left[ \left( \bar{\sigma}_{AB}\right)^{-1} \bar{Y}_{1} \right]}{1-\Tr \left[ \left( \bar{\sigma}_{AB}\right)^{-1} \bar{Y}_{1} \right]}  \ \ \ ,
\end{align}
where the hyperbolic properties of $\bar{\sigma}_{AB} = D_{AB,cosh} \oplus D_{AB,cosh}$ and the relation of Eq.~\eqref{eq:matrixmani} have been used. The physicality relation of $\bar{\sigma}_{AB} - \bar{Y}_1$ leads to,
\begin{equation}
    \det \left( \bar{\sigma}_{AB}^{} - \bar{Y}_{1} + i \Omega \right) =  \sqrt{\det \left( \left( \bar{\sigma}_{AB}\right)^2 - \mathbb{I} \right)} \left( 1 -\Tr \left[ \left( \bar{\sigma}_{AB}^{} - \left( \bar{\sigma}_{AB}\right)^{-1} \right)^{-1} \bar{Y}_{1} \right]  \right) \geq 0 \ \ \ .
    \label{eq:phsic}
\end{equation}
According to Weyl's inequality~\cite{hornmatrixana} and its generalization to symplectic eigenvalues~\cite{Giedkepurestatetrans}, to preserve the physicality of a CM, any normal modes that result in $\det \left( \bar{\sigma}_{AB}^{2} - \mathbb{I} \right) = 0$ must correspond to the kernel of $\bar{Y}_{1}$. Removing the kernel of $\bar{Y}_{1}$ thus ensures the positivity of $\det \left( \left( \bar{\sigma}_{AB}\right)^2 - \mathbb{I} \right)$, while retaining the trace in Eq.~\eqref{eq:phsic}. This leads to $1 -\Tr \left[ \left( \bar{\sigma}_{AB}^{} - \left( \bar{\sigma}_{AB}\right)^{-1} \right)^{-1} \bar{Y}_{1} \right] \geq 0$, hence the domain of Eq.~\eqref{eq:domainkey} is,
\begin{equation}
1 \geq \bar{v} \langle \bar{p}_{0} | \left(D_{C,cosh}^{-1} \oplus D_{C,cosh}^{-1} \right) | \bar{p}_{0} \rangle \geq 0   \ \ \ ,
\label{eq:sqrtreal}
\end{equation}
where the lower bound is a result from CM positivity. Therefore, $|\bar{p}\rangle$ of Eq.~\eqref{eq:pveccons} provides a systematic method of satisfying both Eq.~\eqref{eq:conn1} and Eq.~\eqref{eq:conn2}. With the corresponding projective measurement and classical communication, any physical rank-one Gaussian noise from the $AB$ reduced CM can thus be removed. When the  noise subtraction has the feature of generating a purity ($\nu = 1$) in the symplectic spectrum, i.e., for $P_\sigma$ the number of 1's in the non-degenerate symplectic spectrum $P_{\bar{\sigma}_{AB} '} - P_{ \bar{\sigma}_{AB} } = 1$, no auxiliary mode is needed in the $C$-space measurement protocol such that $|\bar{p}\rangle = |\bar{p}_0\rangle$.

As discussed in Eqs.~\eqref{eq:commuteMandT0}--\eqref{eq:commuteMandT1}, the operator product $S_{\langle p|} = S_{\langle \bar{p}|}\bar{S}_C$ thus aligns the original (lattice) basis in preparation for $Y_1$ noise subtraction via $\phi$-basis projective measurement on the last site in $C$.

\subsection{Constructing projective measurements for any Gaussian noise decomposition}
\label{sec:appasec4}

\begin{figure}[t!]
\centering
\includegraphics[width=0.9\textwidth]{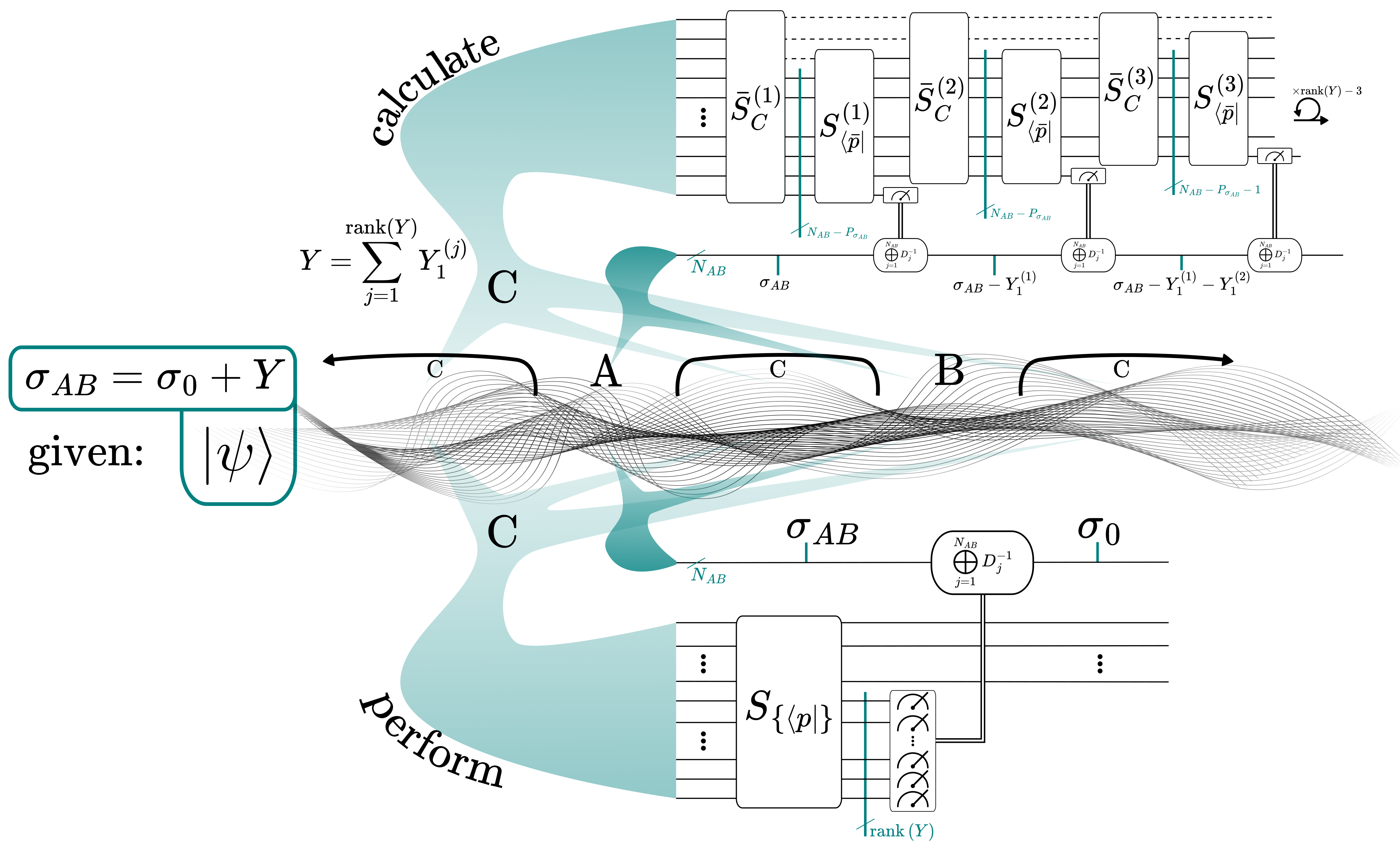}
\caption{Schematic representation of the projective measurement procedures in $C$ realizing Gaussian noise decomposition $\sigma_{AB} = \sigma_{0} + Y$, with classical communication of the measurement results that control local displacement operators in $AB$. The top part of the diagram presents the iterative procedure that calculates profiles of projective measurements for each rank-one noise component, whereas the bottom part depicts an equivalent circuit simplification performing a single transformation in $C$. Since symplectic transformations are unitary in Hilbert space, the symplectic operations updating the covariance matrix, i.e., $\sigma \rightarrow S \sigma S^T$, are depicted in the form of unitary quantum circuitry. Each line of the quantum circuit represents a Gaussian continuous variable mode, where dashed lines indicate pure optical-vacuum modes that have decoupled after transformation into the local normal mode basis. The teal vertical lines with slash notation indicate the number of modes. In the presented scenario, $Y_1^{(1)}$ and $Y_1^{(3)}$ do not create new purities, necessitating a pure optical-vacuum mode in $S_{\langle\bar{p}|}^{(1)}$ and $S_{\langle\bar{p}|}^{(3)}$ as in Eq.~\eqref{eq:pveccons}, thus the number of impurities remains unchanged in the next iteration. However, $Y_1^{(2)}$ is chosen to generate a purity in $\sigma_{AB}$, thus a pure optical-vacuum mode is not utilized in $S_{\langle\bar{p}|}^{(2)}$ as in Eq.~\eqref{eq:pveccons0}.}
\label{fig:circuitdiagram}
\end{figure}

Layering the calculation of a single projective measurement governed by Eq.~\eqref{eq:pveccons} that removes any given Gaussian noise component along with the techniques of Eqs.~\eqref{eq:a42} and~\eqref{eq:rfirstmoments} to update the global pure state shared by $ABC_1$ after each $C_2$ measurement and classical communication to $AB$, a set of projective measurements can be systematically constructed corresponding to any Gaussian noise decomposition. As depicted in Fig.~\ref{fig:circuitdiagram}, this sequence of measurements can be compiled into a single symplectic transformation in $C$ followed by projectively measuring the last $\text{rank}(Y)$ modes of $C$ in the $\phi$ basis. Throughout this appendix, the removal of a rank-one component is assumed to decrease the rank of $Y$ by one (though present techniques remain applicable outside this perspective).

Given an $ABC$ pure state and a noise decomposition $\sigma_{AB} = \sigma_0 + Y$, the following is a summary of the steps developed above for iteratively calculating a set of collective measurements in $C$ whose results, when shared with $AB$, enable the reduced CM of $AB$ to remove the desired set of rank-one noise components:

\begin{enumerate} 

\item Calculate the global pure state with the basis of local normal modes in the form of positively squeezed TMSVS pairs, Eqs.~\eqref{eq:commuteMandT0} and~\eqref{eq:normalform1}, via symplectic operation $\bar{S}_{AB} \oplus \bar{S}_{C}$

\item Select a rank one component of $Y$, e.g., from its eigendecomposition as in Appendix~\ref{sec:appcsec4} or via MNF techniques as in Appendix~\ref{sec:appbsec3}, and denote it as $Y_{1}$. Compute its form in the basis of local normal modes, ${\bar{Y}}_{1} = {\bar{S}}_{AB} Y_{1} {\bar{S}}_{AB}^{T}$.

\item Calculate the single non-zero eigenvalue and corresponding normalized eigenvector of the inverse matrix subtraction, $\left(\bar{\sigma}_{AB} - \bar{Y}_1 \right)^{-1} - \left( \bar{\sigma}_{AB} \right)^{-1} = \bar{v} |\bar{v}\rangle\langle \bar{v}|$.

\item Construct a profile vector, $\langle \bar{p}|$, following Eqs.~\eqref{eq:pveccons0} and~\eqref{eq:pveccons} and build a $C$-space symplectic transformation $S_{\langle \bar{p}|}$ that places the profile in the $\phi$-component of the last mode. Perform $S_{\langle p | } \equiv S_{\langle \bar{p}|} \bar{S}_C$ in the volume.

\item Utilizing Eq.~\eqref{eq:a42}, calculate the global pure state shared by $ABC_1$ after measurement of the last mode in $C$ and classical-communication-controlled displacements in $AB$ governed by (negative) Eq.~\eqref{eq:rfirstmoments}. This process leaves $AB$ with a reduced CM of $\sigma_{AB}-Y_1$.

\item Iterate steps 1--5 on another rank-one component of $Y$ until all components are subtracted.

\end{enumerate}
Through a single iteration of steps 1--5 described above, two symplectic transformations, ${\bar{S}}_{C}^{(1)}$ and $S_{\langle \bar{p}|}^{(1)}$, are calculated in the volume. The second iteration calculates ${\bar{S}}_{C}^{(2)}$ and $S_{\langle \bar{p}|}^{(2)}$, which have smaller matrix dimensions to accommodate the measurement in the previous iteration. Define the overall symplectic transformation, $S_{\left\{ \langle p|\right\}} \stackrel{\text{reorder}}{=} \prod_{i=1}^{\rank \left(Y\right)} \left( \left( S_{\langle \bar{p}|}^{(i)} \oplus \mathbb{I}_{2i-2} \right) \left( \bar{S}_{C}^{(i)} \oplus \mathbb{I}_{2i-2} \right) \right)$, where $\mathbb{I}_{2i-2}$ denotes the identity matrix with matrix dimension $2i-2$, and reorder refers to rearranging the canonical operators from a $\phi\pi$-interleaved structure to one in which all $\phi$ components are placed before the $\pi$ components. To remove $Y$ from $\sigma_{AB}$, it is thus sufficient to apply this single symplectic operation in the $C$ Hilbert space, followed by single-mode displacements in $AB$ governed by the measurement results of the last $\text{rank}(Y)$ $C$-modes projected in the $\phi$-basis.

The number of auxiliary modes in $C$ required by Eq.~\eqref{eq:pveccons} to be introduced (unless provided by modes that decouple in the local normal mode basis as depicted in Fig.~\ref{fig:circuitdiagram}) is upper bounded by $2N_{AB}-P_{\sigma_{AB}} - P_{\sigma_{0}}$ and lower bounded by zero. To see this, note,
\begin{equation}
2N_{AB}-2P_{\sigma_{AB}} \geq \rank \left( Y \right)    \geq \rank \left( \sigma_{AB} + i \Omega \right) - \rank\left( \sigma_{0} +i\Omega \right) = \left( 2N_{AB}-P_{\sigma_{AB}}\right)- \left(2N_{AB}-P_{\sigma_{0}} \right)  ,
\label{eq:ranks}
\end{equation}
where the first inequality comes from Weyl's inequality as discussed in Appendix~\ref{sec:appasec3}, the second inequality leverages the subadditivity of the rank, and the equality of the right side can be seen through transforming ${\sigma + i \Omega}$ with the symplectically diagonalizing (to Williamson normal form) transformation. When $Y$ saturates the lower bound, 
\begin{equation}
    \rank \left( Y \right) = P_{\sigma_0} - P_{\sigma_{AB}} =\rank \left( Y - Y_{1} \right) +1 \geq P_{\sigma_0} - P_{\sigma_{AB}-Y_{1}} +1    \ \ \ ,
\end{equation}
such that $P_{\sigma_{AB}-Y_1} - P_{\sigma_{AB}} \geq 1$, and because subtraction of a rank-one noise component can maximally produce a single additional purity,
\begin{equation}
    \rank \left( \sigma_{AB} - Y_{1} + i \Omega \right) + \rank \left(  Y_{1}  \right) =  2N_{AB}-P_{\sigma_{AB}-Y_{1}} + 1 \geq \rank \left(  \sigma_{AB} + i \Omega \right) = 2N_{AB}-P_{\sigma_{AB}}    \ \ \ ,
\end{equation}
such that $P_{\sigma_{AB}-Y_1} - P_{\sigma_{AB}} \leq 1$, then in this case $P_{\sigma_{AB}-Y_1} - P_{\sigma_{AB}} = 1$, i.e., subtraction of $Y_1$ generates a purity component. Therefore, any $Y$ with $\rank (Y) = P_{\sigma_0} - P_{\sigma_{AB}}$ will not require any auxiliary modes (Eq.~\eqref{eq:pveccons0} will always be sufficient) when performing sequential projective measurement following its  decomposition. Examples of such states are those living on the rim discussed in Appendix~\ref{sec:appcsec1}, as well as those obtained through maximizing/minimizing inter-$DCV$ distance $\xi$ in the finite-volume scalar field vacuum in Appendix~\ref{sec:appcsec3} and~\ref{sec:appcsec4}. The difference between the upper and lower bound of Eq.~\eqref{eq:ranks} is thus an upper bound on the number of  components that require an auxiliary mode in the measurement protocol.

\section{Generalized MNF procedure}
\label{sec:appb}

Understanding the structure of the negativity-contributing subspace ($\mathcal{V_{\mathcal{N}}}$) and its complement ($\mathcal{V}_{\Nslash}$) has led to the identification of local symplectic orthogonality (SOL) in $\mathcal{V_{\mathcal{N}}}$ ($\mathcal{V_{\mathcal{N}}}$-SOL, previously denoted as $\mathcal{N}$-SOL in Ref.~\cite{gao2024partialtransposeguided}) as a necessary and sufficient condition for consolidating many-body negativity, via local operations $S_A \oplus S_B$, into a tensor-product series of $(1_A \times 1_B)$ two-mode entangled pairs~\cite{gao2024partialtransposeguided}. Through a Gaussian classical mixing channel~\cite{WolfGEOF,GiedkeEOF}, negativity consolidation has provided the foundation for a minimum noise filtering procedure ($\mathcal{V_{\mathcal{N}}}$-SOL MNF) that is necessary and sufficient to (constructively) identify how to isolate noise to $\mathcal{V}_{\Nslash}$. Furthermode, when this can be done without expansion of $\mathcal{V}_\mathcal{N}$ (i.e., $\sigma_{AB}$ is a member of the $\mathcal{N}$IC entanglement class), the minimum negativity of the underlying pure state $\sigma_0$ saturates the lower bound set by the mixed-state negativity~\cite{gao2024partialtransposeguided}. In this $\mathcal{N}$IC subspace, not only the negativity, but the entirety of the entanglement may be consolidated into a tensor-product series of $\left(1_A \times 1_B\right)$ two-mode entangled pairs~\cite{gao2024partialtransposeguided}. Beyond $\mathcal{V_{\mathcal{N}}}$-SOL MNF, this appendix introduces the more general PT-SOL and shows that it is necessary and sufficient to identify noise isolated to a SOL-complementary space. 

\subsection{Gaussian classical mixing channel for symmetric two-mode Gaussian states}
\label{sec:appbsec1}

Utilizing the eigendecomposition of $\sigma_{\phi}-\sigma_{\pi}^{-1}$, this section provides an approach to identifying noise isolated to $\mathcal{V}_{\Nslash}$ or $\mathcal{V}_{\mathcal{N}}$ for two-mode symmetric Gaussian states. The underlying pure states revealed by removal of these noise contributions are shown to optimize the GEOF and GEOA, respectively.

First, consider the normal form of symmetric two-mode Gaussian states $\sigma_{2sym} = 
\begin{pmatrix}
a & c_+ \\ 
c_+ & a
\end{pmatrix} \oplus
\begin{pmatrix}
a & c_- \\ 
c_- & a
\end{pmatrix}$, i.e., the case $a = b$ in Eq.~\eqref{eq:twomodeGparam} with $c_{+} \geq |c_{-}|$~\cite{Simonreflection,Duanlocaltrans}. The result of this section is that optimal pure Gaussian states $\sigma_{0,GEOF}= X_{GEOF} \oplus X_{GEOF}^{-1}$ and $\sigma_{0,GEOA}=X_{GEOA} \oplus X_{GEOA}^{-1}$ can be written with eigendecomposition $\sigma_{\phi}-\sigma_{\pi}^{-1}\equiv e_{+}\ket{e_{+}}\bra{e_{+}} + e_{-}\ket{e_{-}}\bra{e_{-}}$, such that $X_{GEOF} = \sigma_{\pi}^{-1} + e_{-}\ket{e_{-}}\bra{e_{-}}$ and $X_{GEOA} = \sigma_{\pi}^{-1} + e_{+}\ket{e_{+}}\bra{e_{+}}$, where the $\pm$ subscript indicates the AB-parity of the associated eigenvector. Explicitly,
\begin{equation}
    \scalemath{0.96}{X_{GEOF} =  \sigma_{\pi}^{-1} + \left( \frac{a-c_{+}}{2}-\frac{1}{2(a-c_{-})}\right) \begin{pmatrix}
1 & -1 \\
-1 & 1 \\
\end{pmatrix}, \quad X_{GEOA} = \sigma_{\pi}^{-1} + \left( \frac{a+c_{+}}{2}-\frac{1}{2(a+c_{-})}\right) \begin{pmatrix}
1 & 1 \\
1 & 1 \\
\end{pmatrix}} \ \ \ .
\label{eq:optsym}
\end{equation}
The positive semidefinite (PSD) noise $Y=\sigma_{2sym}-\sigma_{0}$ in each case reads,
\begin{equation}
    Y_{GEOF} = \begin{pmatrix}
m_{11} & m_{11} & 0 & 0  \\
m_{11} & m_{11} & 0 & 0  \\
0 & 0 & m_{22} & -m_{22}  \\
0 & 0 & -m_{22} & m_{22}   \\
\end{pmatrix}\geq 0,\quad Y_{GEOA} =  \begin{pmatrix}
n_{11} & -n_{11} & 0 & 0  \\
-n_{11} & n_{11} & 0 & 0  \\
0 & 0 & n_{22} & n_{22}  \\
0 & 0 & n_{22} & n_{22}   \\
\end{pmatrix} \geq 0 \ \ \ ,
\label{eq:Yoptsym}
\end{equation}
where $2(a+c_{-})m_{11} = 2(a+c_{+})n_{22} = (a+c_+)(a+c_-)-1$ and $2(a-c_{+})m_{22} = 2(a-c_{-})n_{11}= (a-c_+)(a-c_-)-1$. In the basis of partially transposed normal modes,
\begin{equation}
\scalemath{0.89}{    \Tilde{D} = \Tilde{S}_{2sym} \Tilde{\sigma}_{2sym} \Tilde{S}_{2sym}^{T} = \begin{pmatrix}
\Tilde{\nu}_{+} & 0 & 0 & 0  \\
0 & \Tilde{\nu}_{-} & 0 & 0  \\
0 & 0 & \Tilde{\nu}_{+} & 0  \\
0 & 0 & 0 & \Tilde{\nu}_{-}   \\
\end{pmatrix},\quad \Tilde{S}_{2sym} = \begin{pmatrix}
(\frac{a-c_{-}}{4a+4c_{+}})^{1/4} & (\frac{a-c_{-}}{4a+4c_{+}})^{1/4} & 0 & 0 \\
-(\frac{a+c_{-}}{4a-4c_{+}})^{1/4} & (\frac{a+c_{-}}{4a-4c_{+}})^{1/4} & 0 & 0   \\
0 & 0 & (\frac{a+c_{+}}{4a-4c_{-}})^{1/4} & (\frac{a+c_{+}}{4a-4c_{-}})^{1/4} \\
0 & 0 & -(\frac{a-c_{+}}{4a+4c_{-}})^{1/4} & (\frac{a-c_{+}}{4a+4c_{-}})^{1/4}   \\
\end{pmatrix}} ,
\label{eq:Stildefor2sym}
\end{equation}
where $\Tilde{\nu}_{+}=\sqrt{(a-c_{-})(a+c_{+})} \geq 1$ and $\Tilde{\nu}_{+} \geq \Tilde{\nu}_{-}=\sqrt{(a+c_{-})(a-c_{+})}$. It can be seen that $\tilde{S}\Lambda Y_{GEOF}\Lambda\tilde{S}^T$ has support only in $\mathcal{V}_{\Nslash}$ and $\tilde{S}\Lambda Y_{GEOA}\Lambda\tilde{S}^T$ has support only in $\mathcal{V}_{\mathcal{N}}$, as anticipated by the monotonic relation between classical noise and entanglement. 

In order to see the optimality of Eq.~\eqref{eq:optsym}, note that $\sigma_{2sym}$ is entangled if and only if $\tilde{\nu}_{-} < 1$~\cite{Simonreflection,Duanlocaltrans} and that the extremal solutions for GEOF and GEOA in the two-mode context may be governed by extremizing the negativity of $\sigma_0$. For a valid Gaussian decomposition satisfying $\sigma_{2sym} \geq \sigma_{0}$, denote the PT symplectic eigenvalues of $\sigma_0$ as $\left( \Tilde{\nu},1/\Tilde{\nu},\Tilde{\nu},1/\Tilde{\nu} \right)$ such that $\Tilde{\nu} \geq 1 \geq 1/\Tilde{\nu}$. Since $\Tilde{\nu}_{+} \geq \Tilde{\nu}_{-}$, Weyl's inequality~\cite{hornmatrixana} and its generalization to symplectic eigenvalues~\cite{Giedkepurestatetrans} in PT space yields combined bounds $\Tilde{\nu}_{-} \geq 1/\Tilde{\nu} \geq 1/\Tilde{\nu}_{+}$. As such, when $\sigma_{2sym}$ is entangled, PT symplectic eigenvalues of $\sigma_0$ reading $\left( 1/\Tilde{\nu}_{-},\Tilde{\nu}_{-},1/\Tilde{\nu}_{-},\Tilde{\nu}_{-} \right)$ saturate the upper bound, $\tilde{\nu}_{-}$, providing a solution to GEOF modified from $\sigma_{2sym}$ only in the $\mathcal{V}_{\Nslash}$ subspace. In the alternate case of maximizing the pure-state entanglement, PT symplectic eigenvalues of $\sigma_0$ reading $\left( \Tilde{\nu}_{+},1/\Tilde{\nu}_{+},\Tilde{\nu}_{+},1/\Tilde{\nu}_{+} \right)$ saturate the lower bound, $1/\tilde{\nu}_+$,  providing a (first) solution to GEOA modified from $\sigma_{2sym}$ only in the $\mathcal{V}_{\mathcal{N}}$ subspace.  

The above results can be immediately generalized to $\phi\pi$-uncorrelated two-mode symmetric states without initializing to normal form. This can be seen in that a single-mode squeezing transformation applied to both the A and B mode, $S_{\lambda} =\diag\left( \lambda,\lambda^{-1} \right)$, introduces a common factor of $\lambda^2$ on $\sigma_\phi$ and $\sigma_{\pi}^{-1}$, which maintains the structures of Eqs.~\eqref{eq:optsym} and~\eqref{eq:Yoptsym}.

\subsection{PT-SOL MNF and pure state identification}
\label{sec:appbsec2}

This section extends the techniques of Ref.~\cite{gao2024partialtransposeguided} to consider SOL in any subset of $\tilde{S}$ row vectors, i.e., row vectors of the symplectic operation that diagonalizes the PT CM. Consistent with previous terminology~\cite{NKcorehalo,gao2024partialtransposeguided}, each $(1_A \times 1_B)$ TMSVS space, governed by the corresponding vectors in the PT-SOL subspace, will be referred to as a PT-core pair. By crafting a universal noise parameterization of the PT-core, the proof for PT-SOL MNF is shown below to follow the same procedure for $\mathcal{V_{\mathcal{N}}}$-SOL MNF established in Ref.~\cite{gao2024partialtransposeguided}. 

Towards the PT-SOL generalization, consider first the consolidation transformation for two-mode symmetric gaussian states. From the expression of $\Tilde{S}_{2sym}$ in Eq.~\eqref{eq:Stildefor2sym}, which exhibits SOL in both the $\mathcal{V}_\mathcal{N}$ and $\mathcal{V}_\Nslash$ subspaces, PT-SOL for symmetric two-mode Gaussian states is entirely captured by the combination of $\mathcal{V}_\mathcal{N}$-SOL and $\mathcal{V}_\Nslash$-SOL. Following Eq.~(12) of Ref.~\cite{gao2024partialtransposeguided}, pairs of single-mode squeezing transformations with $\lambda_{A} = \lambda_{B} = \left[(a+c_{-})/(a-c_{+})\right]^{1/4}$ and $-\lambda_{A} = \lambda_{B} = \left[(a-c_{-})/(a+c_{+})\right]^{1/4}$ produce alignment with a TMSVS in the $\mathcal{V}_{\mathcal{N}}$ and $\mathcal{V}_{\Nslash}$ subspaces, respectively, i.e., the alignment that has governed entanglement consolidation in higher-body systems. For the latter, the minus sign can be interpreted as a single-mode $\pi$-phase shift $S_\pi = - \mathbb{I}_2$ that has been included in the $A$-space. This results in a negatively squeezed TMSVS, transforming the matrix structure of $Y_{GEOA}$ to match that of $Y_{GEOF}$, i.e., a universal noise parametrization for the PT-core. 

In multimode contexts, analogous to Eq.~(14) of Ref.~\cite{gao2024partialtransposeguided}, consolidation transformations $S_{A}$ and $S_{B}$ can be designed from the SOL subspace through a symplectic Gram-Schmidt ($\text{sGS}$) procedure as,
\begin{equation}
    S_{A} \oplus S_{B} = \Lambda (S_\pi\ \text{sGS}[\Tilde{S}_{\mathcal{V}_{SOL},A}] \oplus \text{sGS}[\Tilde{S}_{\mathcal{V}_{SOL},B}]) \Lambda \ \ \ ,
    \label{eq:gseq}
\end{equation}
where each $\text{sGS}$ is initiated with the half-space $\tilde{S}$ row vectors in the SOL subspace, $\mathcal{V}_{SOL}$, and completed with an arbitrary spanning basis to produce a symplectic transformation. To smoothly utilize previously established MNF procedures, the $S_\pi$ in Eq.~\eqref{eq:gseq} represents the set of single-mode $\pi$-phase shifts necessary to make the post-consolidation $\tilde{S}$ row vectors in $\mathcal{V}_{SOL}$ AB anti-symmetric.  

Generalizing the above form of $Y_{GEOF}$ for the presence of $\phi\pi$-correlations, Ref.~\cite{gao2024partialtransposeguided} reports the parameterization of PT-core noise isolated to $\mathcal{V}_{\Nslash}$ as,
\begin{equation}
    Y_{c_{1}}=\frac{1}{2}\begin{pmatrix}
y_{11} & y_{11} & y_{12} & -y_{12}  \\
y_{11} & y_{11} & y_{12} & -y_{12}  \\
y_{12} & y_{12} & y_{22} & -y_{22}  \\
-y_{12} & -y_{12} & -y_{22} & y_{22}   \\
\end{pmatrix} \geq 0
\label{eq:Y11cform} \ \ \ .
\end{equation}
Note that this result can also be appreciated in the present context as maintaining PT-SOL subspace alignment between the mixed state and a TMSVS~\footnote{The following determines the general form of noise that preserves alignment with a TMSVS in the PT-SOL subspace of a $\phi\pi$-correlated mixed Gaussian state, $\sigma''$. Under local operations, such a state can be transformed into a $\phi\pi$-uncorrelated one~\cite{Simonreflection,Duanlocaltrans} without modifying the SOL property~\cite{gao2024partialtransposeguided}. Upon symbolic calculations, it is further seen that for such two-mode Gaussian states in normal form~\cite{Simonreflection,Duanlocaltrans}, $AB$ symmetry is necessary and sufficient for $SOL$, which is present in both the $\mathcal{V}_{\mathcal{N}}$ and $\mathcal{V}_{\Nslash}$ subspaces. Therefore, the symplectic transformation $\tilde{S}''$ that diagonalizes the PT CM, $\Lambda \sigma'' \Lambda$, must locally connect to $\tilde{S}'$ that diagonalizes the PT consolidated $\sigma_{2sym}$ through $\tilde{S}'' = \tilde{S}' \Lambda(S_A \oplus S_B)^{-1} \Lambda$~\cite{gao2024partialtransposeguided}. In order to align with a TMSVS in the SOL subspace, $\Tilde{S}^{''}$ must preserve the post-consolidation structure of $\Tilde{S}^{'}$. Under local operations that preserve the normal mode decomposition of the PT CM, $(O_{A}\oplus O_{B})\Tilde{S}^{''}$, preserving the post-consolidation structure is achieved by $S_{A} =S_{B}^{T}=O_{B}$ for consolidation in $\mathcal{V_{\mathcal{N}}}$ and $S_{A} =S_{B}^{T}=O_{A}$ for consolidation in $\mathcal{V_{\Nslash}}$. This updates Eq.~\eqref{eq:Yoptsym} to include $\phi\pi$-correlations. Therefore, parameterization of noise for the PT-core arrives at Eq.~\eqref{eq:Y11cform} from a perspective complementary to that utilized in Ref.~\cite{gao2024partialtransposeguided}.}. Consistent with the $S_{\pi}$-generated universality discussed above, identification of noise in the form of Eq.~\eqref{eq:Y11cform} for a $\mathcal{V}_{\Nslash}$-SOL consolidated core (i.e., noise isolated to $\mathcal{V}_{\mathcal{N}}$ for GEOA applications) will result in a negative-squeezed TMSVS.

Denote the off-diagonal block of the CM between the first PT-core and the rest of the system $\sigma_{c_{1} r}$. With the universal noise parameterization established above, column vectors of $\sigma_{c_1 r}$ exhibit the same form, $\ket{l}\equiv \left( l_{i},l_{i},-l_{j},l_{j}\right)^{T}$, as that of $\mathcal{V}_{\mathcal{N}}$-SOL consolidation in Ref.~\cite{gao2024partialtransposeguided}. Because the proof for $\mathcal{V_{\mathcal{N}}}$-SOL MNF only relies on the explicit parametrization of $Y_{c_{1}}$ and $\ket{l}$, that for PT-SOL MNF as constructed follows the same procedure as developed in Ref.~\cite{gao2024partialtransposeguided}. In order to both retain a physical CM and explicitly produce separability between the core and the rest of the system upon identification of $Y$, $Y_{c_{1} r} = \sigma_{c_{1} r}$ is removed in the CM off diagonal block between the PT-core and rest of the system, and the saturation of $Y_{r}=Y_{c_{1} r}^{T}Y_{c_{1}}^{-1}Y_{c_{1} r}$ will be used to define one iteration of the MNF process. Due to Eq.~(21) of Ref.~\cite{gao2024partialtransposeguided}, this noise form does not affect the SOL subspace of the remaining CM, thus ensuring that the PT-SOL MNF procedure generates a pure state with all noise isolated in the complementary subspace. 

Conversely, every pure state decomposition with noise isolated to the SOL-complementary space can be generated from the state identified by the PT-SOL MNF procedure. To see this via Weyl's inequality~\cite{hornmatrixana,Giedkepurestatetrans}, adding noise on a pure Gaussian state isolated to the SOL-complementary  space will retain alignment, and hence SOL, in the noiseless subspace~\cite{gao2024partialtransposeguided}. PT-SOL MNF enforces this alignment in the SOL-subspace between a mixed state and underlying pure state, subtracting the minimum noise in the complementary space to support this alignment, which results in noise isolated to its SOL-complementary space. Therefore, PT-SOL is necessary and sufficient to identify pure Gaussian states with Gaussian noise isolated to the SOL-complementary space.

\subsection{\texorpdfstring{$\mathcal{V}_{\pm}$}{Vpm} PT-SOL MNF and assisted measurement profiles}
\label{sec:appbsec3}

\begin{figure*}
\centering
\includegraphics[width=1\textwidth]{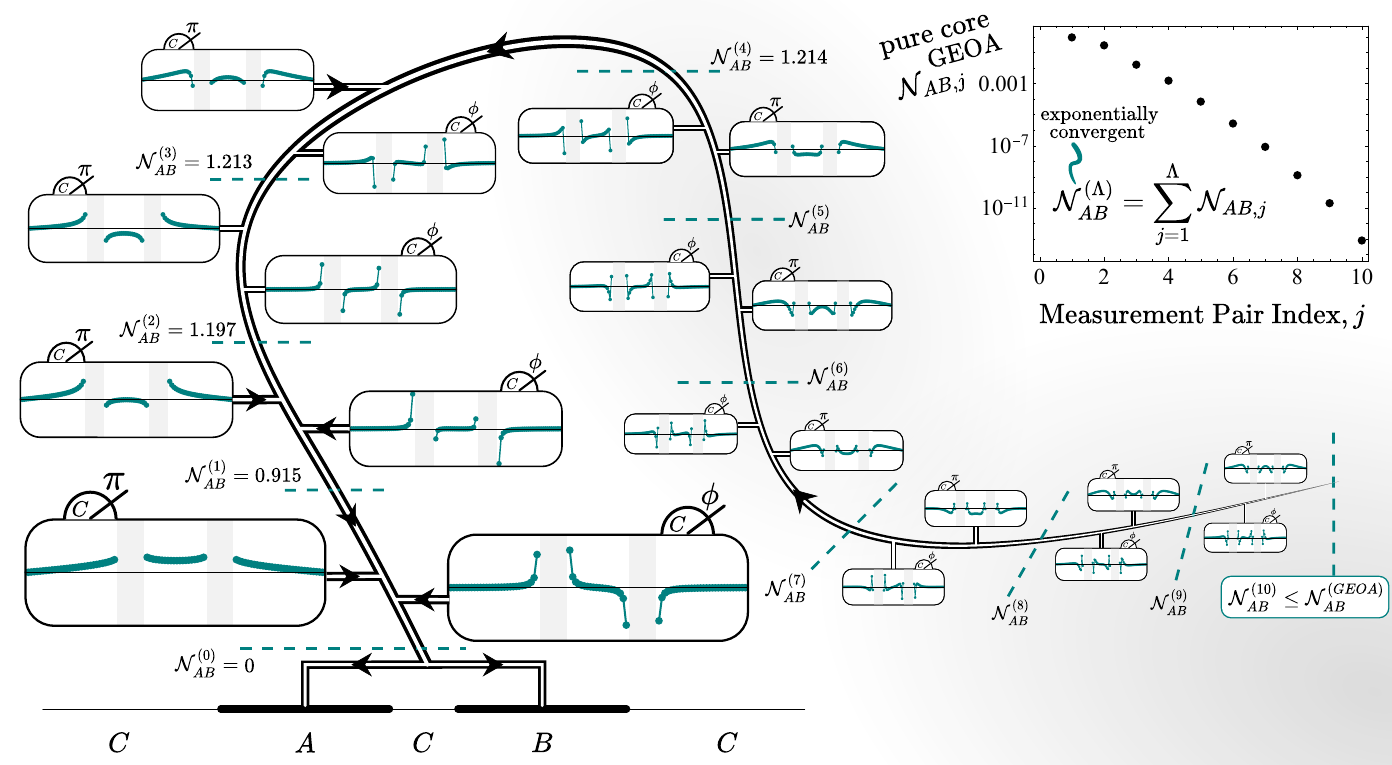}
  \caption{Classical communication procedure for projective measurements in $C$ that produces the $AB$ pure state lower-bounding the GEOA between disjoint regions of the 100-site OBC lattice scalar field vacuum, $(m, d, \tilde{r}) = (0,10, 20)$. After the measurement and communication of each $\phi$-$\pi$ collective profile pair, $j$, an additional $(1_A \times 1_B)$ pure core is generated in $AB$ with negativity $\mathcal{N}_{AB,j}$. Governing these profiles by the rank-one matrix decomposition provided by the $\mathcal{V}_{\pm}$-SOL MNF procedure allows maximization of $\mathcal{N}_{AB,1}$. As depicted by the diminishing size of profiles at later points in the measurement and communication sequence, and shown explicitly in the associated plot, the total pure negativity available in $AB$ after $\Lambda$ pairs, $\mathcal{N}_{AB}^{(\Lambda)} = \sum_{j = 1}^\Lambda \mathcal{N}_{AB,j}$, rapidly converges. A symplectic transformation, $S_{\{ \langle p| \} }$, localizing these profiles within the volume $C$ prior to measurement may be constructed following Eqs.~\eqref{eq:gsreplace1} and~\eqref{eq:gsreplace2}.}
  \label{fig:fig4}
\end{figure*}

Consider a pure two-mode state in $AB$ created through assistance via measurements in $C$ and local symplectic transformations in $AB$. In a pure Gaussian state, the smallest PT symplectic eigenvalue (which governs the the maximum entanglement of a pure two-mode component) is the reciprocal of the largest one. 
In a mixed Gaussian state, applying Weyl's inequality~\cite{hornmatrixana} to the PT symplectic eigenvalues~\cite{Giedkepurestatetrans,gao2024partialtransposeguided} reveals that the largest PT symplectic eigenvalue of an underlying pure state is upper bounded by that of the mixed state. Therefore, the maximum entanglement of a pure two-mode state extracted with the assistance of measurements in $C$ from a multimode mixed state is upper bounded by the inverse of the maximum PT symplectic eigenvalue. 

To demonstrate an assistance process that achieves this optimal entangled pair for disjoint regions of a scalar field vacuum, this section first introduces $\mathcal{V}_{\pm}$, a PT-SOL space containing the maximum PT symplectic eigenvalue. The set of projective measurements in $C$ designed corresponding to the $\mathcal{V_{\pm}}$ PT-SOL MNF not only saturates the upper bound collecting maximum entanglement of a two-mode pure state in the assisted paradigm, but also provides a hierarchical structure of measurements associated with the present GEOA bound. With the establishment of PT-SOL MNF in Appendix~\ref{sec:appbsec2}, the proof for $\mathcal{V_{\pm}}$-SOL MNF becomes straightforward and reduces to proving that $\mathcal{V}_\pm$ is a PT-SOL subspace.

Following Ref.~\cite{gao2024detectingspacelikevacuumentanglement}, for the special case of $AB$ symmetric PT CMs with symmetric $\sigma_{\phi,AB}$ and $\sigma_{\pi,AB}$, and vanishing $\phi$-$\pi$ matrix elements, there exist subspaces $\mathcal{V_{\mp}}$ and $\mathcal{V_{\pm}}$ spanned by symplectic eigenvectors of PT CM block combinations,
\begin{equation}
 \Tilde{\sigma}_{\mp} \equiv \left( \sigma_{\phi,A} - \sigma_{\phi,AB} \right) \oplus \left(\sigma_{\pi,A} + \sigma_{\pi,AB} \right), \quad \Tilde{\sigma}_{\pm} \equiv \left( \sigma_{\phi,A} + \sigma_{\phi,AB} \right) \oplus \left( \sigma_{\pi,A}  -\sigma_{\pi,AB} \right) \ \ \ ,
     \label{eq:pmmp} 
\end{equation}
obtained via applying transformation $S_{0,d} \equiv \frac{1}{\sqrt{2}}\begin{pmatrix} \mathbb{I}_{d} & \mathbb{I}_{d} \\ -\mathbb{I}_{d} & \mathbb{I}_{d} \end{pmatrix}$ to both the $\phi$ and $\pi$ space of the PT CM. Examples of such states include: $\sigma_{2sym}$, the CMs of symmetric disjoint regions of latticed free scalar vacuum, or a Gaussian approximation of local axial motional modes in trapped ion chains~\cite{NKphononatural}. Here, $\mathcal{V_{\mp}}$ and $\mathcal{V_{\pm}}$ are defined across the $AB$ bipartition and collectively form the full $AB$ space. The SOL of $\mathcal{V_{\mp}}$ and $\mathcal{V_{\pm}}$ can be seen from the fact that there always exists symplectic transformations $\tilde{S}_{\mp}$ and $\tilde{S}_{\pm}$ diagonalizing $\tilde{\sigma}_{\mp}$ and $\tilde{\sigma}_{\pm}$~\cite{gao2024detectingspacelikevacuumentanglement}, resulting in complete SOL structure $\langle\tilde{S}_{i,A}|\Omega|\tilde{S}_{j,A}\rangle = \langle\tilde{S}_{i,A}|\Omega|\tilde{S}_{j,A}\rangle = \Omega_{ij}/2 $ for both the $\mathcal{V}_{\pm}$ and $\mathcal{V}_{\mp}$ diagonalizing symplectic operators. Thus, the combined symplectic operator that diagonalizes the CM organized by the $\mathcal{V}_{\pm}$ subspace is,
\begin{equation}
    \tilde{S}  \stackrel{\text{reorder}}{=} (\tilde{S}_{\pm} \oplus \tilde{S}_{\mp} ) S_{0,2d} \ \ \ .
\end{equation}
The consolidation transformation governed by $\mathcal{V_{\pm}}$ can thus be determined from $\tilde{S}$ ($\phi\pi$-interleaved operator ordering),
\begin{equation}
    S_A \oplus S_B = \Lambda\left(-\sqrt{2}\tilde{S}_{\mathcal{V}_{\pm},A} \oplus \sqrt{2}\tilde{S}_{\mathcal{V}_{\pm},B} \right) \Lambda = -\tilde{S}_{\pm} \oplus \tilde{S}_{\pm}  \ \ \ ,
    \label{eq:consopmsol}    
\end{equation}
where $\tilde{S}_{\mathcal{V}_{\pm},A}$ denotes the A half-space of the $\mathcal{V}_{\pm}$ subspace of $\tilde{S}$. For two-mode symmetric states, an alternative way of finding the consolidation transformation is given by calculating symplectic transformations that diagonalize $\Tilde{\sigma}_{\mp,2sym} = \left( a-c_{+} \right) \oplus \left( a+c_{-} \right)$ and $\Tilde{\sigma}_{\pm,2sym}=\left( a+c_{+} \right) \oplus \left( a-c_{-} \right)$. Therefore, it is seen that $\mathcal{V_{\mathcal{N}}}= \mathcal{V_{\mp}}$ and $\mathcal{V}_{\Nslash} = \mathcal{V_{\pm}}$ for $\sigma_{2sym}$. For the many-mode context of disjoint regions within latticed free scalar field vacuum, however, $\mathcal{V_{\mathcal{N}}} \subseteq \mathcal{V_{\mp}}$ and $\mathcal{V_{\Nslash}} \supseteq \mathcal{V_{\pm}}$~\cite{gao2024detectingspacelikevacuumentanglement}. Guiding the optimal GEOA procedure, the largest PT symplectic eigenvalue is included in $\mathcal{V_{\pm}}$, which can be analytically proved with Weyl's inequality through noting $\Tilde{\sigma}_{\pm} \geq \Tilde{\sigma}_{\mp}$ for CMs that share the scalar field property of satisfying $\sigma_{\phi,AB} \geq 0$ and $\sigma_{\pi,AB} \leq 0$, i.e., Eq.~(2) of Ref.~\cite{gao2024detectingspacelikevacuumentanglement}.

For example, in a free scalar field vacuum of 100 bosons with open boundary conditions (OBC), disjoint regions $(d, \tilde{r}, m) = (10,20,0)$ are entangled with logarithmic negativity $\mathcal{N}=0.000635$. After consolidation, the tensor-product set of PT-core pairs enables the negativity calculation of $\mathcal{V_{\pm}}$ PT-SOL MNF pure state identification using Eqs.~\eqref{eq:optsym} for each PT-core pair. The resulting GEOA for PT-cores reads $\mathcal{N} = 0.915, 0.282, 0.0161, 0.00152, 6.82 \times 10^{-5}, 2.66 \times 10^{-6},8.42 \times 10^{-8},1.26 \times 10^{-9},2.02 \times 10^{-11},8.07 \times 10^{-14}$ for a total of $\mathcal{N}_{0}=1.21$. Again due to Eq.~(21) of Ref.~\cite{gao2024partialtransposeguided}, the iterative filtration does not affect the SOL subspace of the remaining CM, hence these GEOA values for PT-cores will persist through the MNF procedure. For each PT-core, the $\mathcal{V_{\pm}}$-SOL MNF procedure yields a rank-two $Y$ that may be further decomposed into a pair of rank-one components isolated respectively to the $\pi$ and $\phi$ phase space. With contributions from all PT-cores, there thus exists a rank-one decomposition of $Y$ guided by $\mathcal{V}_{\pm}$ that achieves the maximal two-mode entanglement (GEOA) after its first two $C$-measurements. Note that this rank-one decomposition is different from the eigendecomposition of $Y$. 

The noise structure plays a crucial role guiding the measurement processes of assisted entanglement. As discussed in Appendix~\ref{sec:appa}, each rank-one component can be subtracted from the disjoint regions CM by performing projective volume measurement and classical communication of the measurement outcome to inform local displacements. With measurements governed by the collective profiles presented in Fig.~\ref{fig:fig4}, the PT-cores are sequentially purified in order of their GEOA contributions. Crucial for the many-mode context toward continuum fields, this sequence provides a hierarchical framework to guide the entanglement purification between spacelike field regions given a finite number of measurements performed in the volume.

As discussed in Sec.~\ref{sec:IIA}, because the $4d$-dimensional noise matrix identified by the $\mathcal{V_{\pm}}$-SOL MNF procedure has rank $2d$, the collective modes for volume measurement can be designed solely from the $2d$ volume normal modes that are entangled with the patches, i.e., by Eq.~\eqref{eq:pveccons0} without Eq.~\eqref{eq:pveccons}. Leveraging the freedoms of single-mode squeezing operations and $S_\pi$ phase shift that leaves the measurement outcome invariant, the profiles presented in Fig.~\ref{fig:fig4} are chosen to be normalized with positive left tail. The remaining freedoms in $S_{\{ \langle p| \} }$ come from measurement-irrelevant rows of each $S_{\langle \bar{p}|}$. This is due to the fact that various completions of $S_{\langle \bar{p}|}$ cannot be connected by a symplectic transformation within the volume. For each $\phi(\pi)$ profile in the creation of Fig.~\ref{fig:fig4}, the pure single normal modes in the volume that decouple from the rest of the system are initially removed. The measurement irrelevant rows in the $\phi(\pi)$ space of each $S_{\langle \bar{p}|}$ are chosen to be an identity with the last(penultimate) row removed, with the symplectic condition completing the alternate phase space of the operator via inverse transpose.

Toward laboratory implementations, given the $\phi\pi$-uncorrelated profiles as shown in Fig.~\ref{fig:fig4}, a symplectic transformation within the scalar field volume ($C$) that places each profile onto a local mode prior to projective measurement (bottom of Fig.~\ref{fig:circuitdiagram}) may be constructed following $S_{\{ \langle p| \} } = \begin{pmatrix} L & 0 \\ 0 & L^{-T} \end{pmatrix}$ with,
\begin{equation}
   L = \begin{pmatrix} | p_{\phi_1} \rangle & .. & | p_{\phi_d} \rangle & | p^{'}_{\pi_1} \rangle & .. & | p^{'}_{\pi_d} \rangle & | p_{\perp_1} \rangle & .. & | p_{\perp_{N_{c}-2d}} \rangle \end{pmatrix}^{T} \ \ \ ,
   \label{eq:gsreplace1}
\end{equation}
and 
\begin{subequations}
\begin{align}
   & \begin{pmatrix} | p^{'}_{\pi_1} \rangle & .. & | p^{'}_{\pi_d} \rangle \end{pmatrix}^{T} = \begin{pmatrix} | p^{}_{\pi_1} \rangle & .. & | p^{}_{\pi_d} \rangle \end{pmatrix}^{-1} \ \ \ , \\ 
   & \begin{pmatrix}
      | p_{\perp_1} \rangle & .. & | p_{\perp_{N_{c}-2d}} \rangle
   \end{pmatrix}^{T} = \nullspace \left( \begin{pmatrix} | p_{\phi_1} \rangle & .. & | p_{\phi_d} \rangle & | p^{'}_{\pi_1} \rangle & .. & | p^{'}_{\pi_d} \rangle \end{pmatrix}^{T} \right) \ \ \ ,
\end{align}
\label{eq:gsreplace2}%
\end{subequations}
where $| p^{}_{\phi_i}\rangle$, $| p^{}_{\pi_i}\rangle$ for $i \in \{1, \ldots, d\}$ denote the $2d$ symplectically orthogonal profiles, $(\cdot)^{-1}$ denotes pseudoinverse and $\nullspace(\cdot)$ denotes a list of vectors spanning the nullspace of the corresponding matrix. Equations~\eqref{eq:gsreplace1} and~\eqref{eq:gsreplace2} provide a computationally advantageous technique for completing $S_{\{\langle p|\}}$ from $2d$ $\phi\pi$-uncorrelated profiles that both maintains the $\phi\pi$-uncorrelated structure and avoids the use of precision-intensive sGS procedures.

\section{Semidefinite conic geometries}
\label{sec:appc}

Analyzing the structure of Gaussian noise via semidefinite cones has enabled the identification of the GEOF for arbitrary two-mode Gaussian states~\cite{WolfGEOF}, with the optimal solution living on the rim of a double-cone volume ($DCV$). Further characterizing entanglement of Gaussian states, the same framework yields the optimal solutions for the GEOA of tensor-product two-mode Gaussian states and GLEMS~\cite{PhysRevA.101.052305}. As such, Appendix~\ref{sec:appcsec1} presents a compact parametrization of the $DCV$ rim, updating the existing literature on its calculation. The optimal solutions derived in Appendix~\ref{sec:appbsec1} are shown to represent specific instances of this general $DCV$ rim parametrization. 

Extending the geometric framework of semidefinite cones to multimode Gaussian states leads to the developments presented in the following subsections. Appendix~\ref{sec:appcsec2} introduces entanglement quantification based on multi-dimensional $DCV$s, and motivates an inter-$DCV$ distance as a necessary and sufficient entanglement quantifier between disjoint regions of scalar field vacuum. The corresponding maximization and minimization of this distance, detailed in Appendix~\ref{sec:appcsec3} and Appendix~\ref{sec:appcsec4}, yield new upper and lower bounds to the underlying pure state entanglement that, in both cases, are parametrically tighter than results obtained by earlier techniques. Notably, the result presented in Appendix~\ref{sec:appbsec3} is found to agree exactly with the maximization of this distance.

\subsection{Two-mode \texorpdfstring{$DCV$}{DCV} rim optimization}
\label{sec:appcsec1}

While reviewing the concept of semidefinite cones, this appendix updates and refines the procedures of GEOF calculation for arbitrary two-mode Gaussian states discussed in Refs.~\cite{WolfGEOF,It3} and provides the optimal underlying pure Gaussian state CM. The explicit calculation of this CM is useful, e.g., in determining the associated collective measurement performed in the volume as discussed in Sec.~\ref{sec:IIA} and Appendix~\ref{sec:appa}, or in preparing mixed Gaussian states as classical mixtures of pure Gaussian states.

Symmetric real two-dimensional matrices can be parametrized as,
\begin{equation}
M  = x \sigma_x+ z \sigma_z+y  \mathbb{I}=\begin{pmatrix}
y+z & x \\
x & y-z \\
\end{pmatrix} \ \ \ ,
\label{eq:generaltwomodeG}
\end{equation}
where $\sigma_x$, $\sigma_z$ and $\mathbb{I}$ are the Pauli x, z and identity matrix. The eigenvalues of M are $y \pm \sqrt{x^2 + z^2}$.  Thus the positive semidefinite (PSD) condition on y is $y_{PSD} \geq \sqrt{x^2 + z^2}$, and the negative semidefinite (NSD) condition on y is $y_{NSD} \leq -\sqrt{x^2 +z^2}$. With this parametrization, sign-definite matrices are contained within a double-cone oriented along the $y$-axis with apex located at $(z, x, y) = \vec{0}$, PSD matrices reside in the upward ($+\hat{y}$, forward light) cone while NSD matrices reside in the downward ($-\hat{y}$, backward light) cone. Points outside (spacelike), for example $(z,x,y) = (1,1,1)$ or $(1,1,-1)$, are neither PSD nor NSD. This double cone is infinite, right, circular, and characterized by a $\pi/2$ opening angle. Furthermore, every matrix $M'$ associated with a point in this 3-dimensional space may serve as the apex of a shifted double-cone with matrices $M$ satisfying $M -M' \geq 0$ and $M-M' \leq 0$ residing in the upward and downward cones, respectively. Matrices $M -M'$ corresponding to points within these cones are rank-two, while those corresponding to points on the surface are rank-one.

Recall that any two-mode Gaussian CM (pure or mixed) can be locally transformed to the following normal form~\cite{Simonreflection,Duanlocaltrans},
\begin{equation}
\sigma_{2}  = \begin{pmatrix}
a & c_{+} \\
c_{+} & b \\
\end{pmatrix} \oplus \begin{pmatrix}
a & c_{-} \\
c_{-} & b \\
\end{pmatrix} \equiv \sigma_{\phi} \oplus \sigma_{\pi} \ \ \ .
\label{eq:twomodeGparam}
\end{equation}
This can be achieved, for example, by first symplectically diagonalizing (to Williamson normal form) each of the reduced CMs individually, then utilizing the singular value decomposition and symplectic Gram-Schmidt to determine single-mode phase shifts that diagonalize the matrix of mode-mode correlation functions. As a transformation by a pair of single-mode operators, the entanglement between the modes is not altered in this procedure. The relationship $c_{+} \geq |c_{-}|$ may be chosen without loss of generality, as the two may be interchanged through a single-mode $\pi/2$ phase shift applied to each mode, and a relative minus sign can be interchanged through a single-mode $\pi$ phase-shift applied to one mode. From the physicality condition, $\sigma_{2} + i \Omega \geq 0$, the CM diagonal blocks respect $\sigma_\phi \geq \sigma_\pi^{-1} >0$.

\begin{figure}[t!]
\centering
\includegraphics[width=0.88\textwidth]{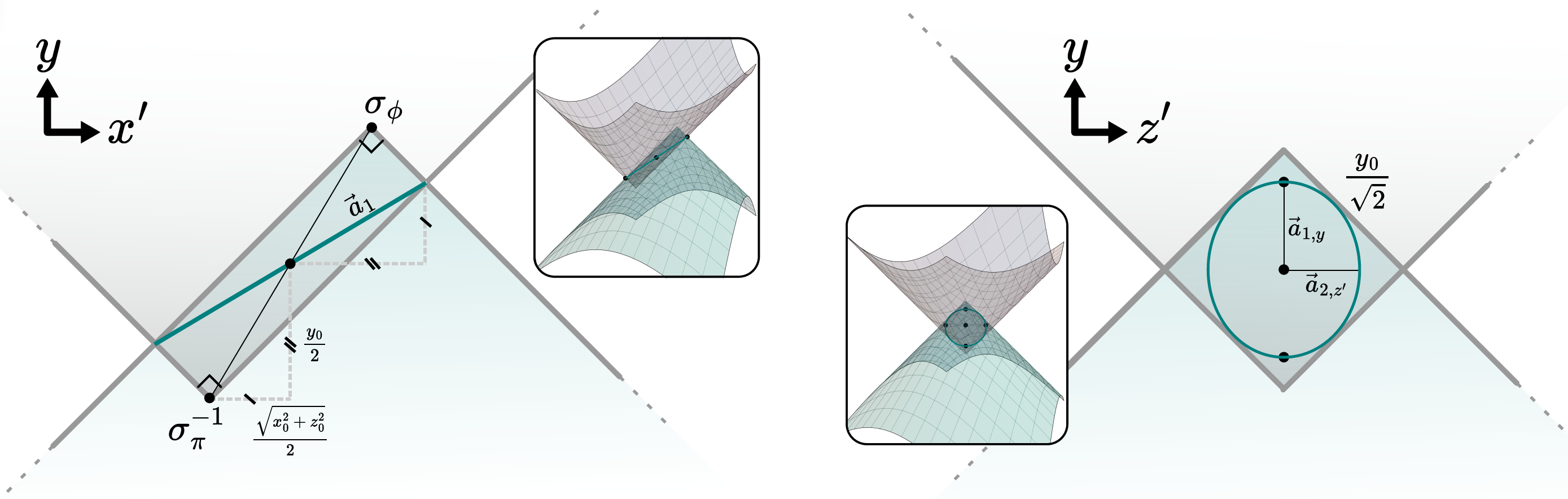}
  \caption{Projections of $DCV$ to (left) plane $x'$-y that contains the semi-major axis of the ellipse and (right) plane $z'$-y that contains the semi-minor axis of the ellipse.}
  \label{fig:dcvGeo_xpyplane}
\end{figure}

An underlying pure state CM $\sigma_0 \equiv \sigma_2-Y$ with minimal entanglement subject to $Y\geq 0$, and thus leading to GEOF of $\sigma_2$, has been previously shown to be available within the $\phi \pi$-uncorrelated subspace, i.e., to also have the block structure of Eq.~\eqref{eq:twomodeGparam}~\cite{WolfGEOF}. With $\sigma_{0}  \equiv X \oplus X^{-1}$, the PD property of $\sigma_\pi$ and $X$ yield combined bounds of,
\begin{equation}
    \sigma_\phi \geq X \geq \sigma_\pi^{-1} \ \ \ .
\end{equation}
Every $X$ satisfying these constraints lives within or on the surface of a double-cone volume (DCV), i.e., the enclosed volume when one cone in a double cone is shifted such that its apex resides within the second cone, as depicted in Fig.~\ref{fig:dcvGeo_xpyplane}. To minimize the pure-state entanglement within the $DCV$, it is convenient to maximize the purity of the reduced density matrix, $1/\sqrt{\det \sigma_1} \equiv 1/\sqrt{m}$, and thus minimize the local determinant $m(X)=1+\frac{X_{AB}^{2}}{\mathrm{Det}\left[X\right]}=1+\frac{(X^{-1})_{12}^{2}}{\mathrm{Det}\left[X^{-1}\right]}$ as it provides a monotonically increasing function of entanglement.

In Ref.~\cite{WolfGEOF}, it is observed that since $X$ is enclosed within the $DCV$, for any $\sigma_\phi > X$ or $X > \sigma_\pi^{-1}$, one can always choose a perturbed matrix, either $\delta_1 = \epsilon \mathbb{I}$ or $\delta_2 = (X^{-1} + \epsilon \mathbb{I})^{-1} - X$ with $\epsilon > 0$, to reduce the underlying pure state entanglement. While $\delta_1$ moves $X$ in the positive $y$-direction until it reaches the downward cone surface (i.e., $\det(\sigma_\phi - X) = 0$), $\delta_2$ further decreases the entanglement and moves $X$ until it reaches the upward cone surface (i.e., $\det(X - \sigma_\pi^{-1}) = 0$). By sending an arbitrary $X$ to the upper and lower $DCV$ surfaces while decreasing the underlying pure state entanglement, the surfaces constitute the subset where optimal $X$ lives. For the case  in which the optimal $X$ is unique, it thus lives on the intersection, i.e., the $DCV$ rim. Updating the argument in Ref.~\cite{WolfGEOF} to allow cases where optimal solutions are not unique, note that $\delta_1$ and $\delta_2$ move $X$ towards the $y$-axis within a plane defined by the initial starting point and the $y$-axis ($X$-$y$-axis plane). By iterating between upward and downward shifts, when the $y$-axis intersects the $DCV$ (i.e., $\text{spec}_{\text{min}}(\sigma_{\phi}) \geq \text{spec}_{\text{max}}(\sigma_{\pi}^{-1})$), which is sufficient for $\sigma_{2}$ to be separable, $X$ converges onto the $y$-axis. Otherwise, it converges onto a point on the rim which is defined by the intersection between the $X$-$y$-axis plane and the DCV. However, as will be shown below, when $\sigma_{2}$ is separable, separable solutions can be analytically confirmed on the rim. Thus, the rim contains a subset where an optimal solution always resides.

Geometrically, the rim of the $DCV$ is an ellipse. The optimal $X$ residing in this ellipse may be parametrized by a single angular variable in terms of the semi-major $\vec{a}_1$ and semi-minor $\vec{a}_2$ axis as
\begin{equation}
X_{GEOF}=\begin{pmatrix}
y_e+z_e & x_e \\
x_e & y_e-z_e \\
\end{pmatrix},\quad
    (z_e,x_e,y_e) = \vec{c} + \vec{a}_1 \cos \theta + \vec{a}_2 \sin \theta \ \ \ ,
\end{equation}
with the point $\vec{v}_{\sigma_\pi^{-1}} = \left( z_{\sigma_\pi^{-1}}, x_{\sigma_\pi^{-1}}, y_{\sigma_\pi^{-1}} \right)$ representing $\sigma_\pi^{-1}$ in the matrix vector space having components,
\begin{equation}
   z_{\sigma_{\pi}^{-1}} = -\frac{a-b}{2\gamma_{c}} , \quad x_{\sigma_{\pi}^{-1}} = - \frac{c_{-}}{\gamma_{c}} , \quad y_{\sigma_{\pi}^{-1}} = \frac{a+b}{2\gamma_{c}}  \ \ \ ,
\end{equation}
where $\gamma_{c} \equiv a b-c_{-}^2 = \det (\sigma_{\pi}) > 0$, the point $\vec{v}_0 = \left( z_0, x_0, y_0\right)$ representing $\sigma_\phi - \sigma_\pi^{-1}$ having components,
\begin{equation}
    z_0 = \frac{a-b}{2}\left(1+\frac{1}{\gamma_{c}}\right) , \quad x_0 = c_{+} + \frac{c_{-}}{\gamma_{c}} , \quad y_0 = \frac{a+b}{2}\left(1-\frac{1}{\gamma_{c}}\right) \ \ \ ,
\end{equation}
and the center of the ellipse located at $ \vec{c} =\vec{v}_{\sigma_\pi^{-1}} + \frac{1}{2} \vec{v}_0$.

In order to evaluate $(z_e,x_e,y_e)$, the introduction of a rotation matrix allows us to leverage the symmetry of the $DCV$, leading to an analytical expression of the ellipse. Define the rotation matrix, 
\begin{equation}
   U_{\alpha}= \begin{pmatrix}
    \cos{\alpha} & -\sin{\alpha} & 0\\
    \sin{\alpha} & \cos{\alpha} & 0 \\
       0 & 0 & 1\\
\end{pmatrix} \ \ \ .
\end{equation}
This rotation describes an active rotation of a vector counter-clockwise about the $y$-axis or new coordinates rotated clockwise such that $U_{-\alpha} \vec{z} = \vec{z}^{'}$, $U_{-\alpha} \vec{x} = \vec{x}^{'}$ and $U_{-\alpha} \vec{y} = \vec{y}$. Shifting $\sigma_\pi^{-1}$ to the origin, the angle $\sin\alpha = z_0/\sqrt{x_0^2+z_0^2}$ and $\cos \alpha = x_0/ \sqrt{x_0^2 + z_0^2}$~\footnote{The angle itself may be expressed as $\alpha = \left( \arcsin \frac{z_0}{\sqrt{x_0^2 + z_0^2}} - \pi\frac{1-\text{sgn}_0\left( x_0\right)}{2} \right) \text{sgn}_0 \left(x_0 \right)$ with $\text{sgn}_0(x)$ the sign function modified at the origin such that $\text{sgn}_0(0) = 1$} aligns the semi-major axis of the elliptical intersection with the $x'$-$y$ plane, where the center of the ellipse is parametrized by $\frac{1}{2} \vec{v}_0' = \frac{1}{2}U_{\alpha}\vec{v}_{0} = \frac{1}{2} \left( 0, \sqrt{x_0^2+z_0^2}, y_0\right)$. With the opening angle of the cone $\pi/2$, as shown in the left panel of Fig.~\ref{fig:dcvGeo_xpyplane}, the semi-major axis vector is parametrized as $ \Vec{a}_{1}^{(z', x', y)}= \frac{1}{2}(0,y_{0},\sqrt{x_{0}^2+z_{0}^2})$. Transforming back to the $z$-$x$-$y$ frame,
\begin{equation}
    \Vec{a}_{1}^{(z, x, y)} =U_{-\alpha}\Vec{a}_{1}^{(z', x', y)}= \frac{1}{2}\left(y_0 \sin{\alpha}, y_0 \cos{\alpha}, \sqrt{x_{0}^{2}+z_{0}^{2}}\right) \ \ \ .
\end{equation}
By symmetry, the semi-minor axis of the ellipse points solely in the $z'$ direction. Projected into the $z'$-$y$ plane, the ellipse is bounded by a square with side length $y_0/\sqrt{2}$, as shown in the right panel of Fig.~\ref{fig:dcvGeo_xpyplane}. The side of this square is related to the axes of the 2D projected ellipse by $y_0/\sqrt{2} = \sqrt{2} \sqrt{\vec{a}_{1,y}^2 + \vec{a}_{2,z'}^2 }$. As such, $ \Vec{a}_{2}^{(z', x', y)} = (|\Vec{a}_{2}|,0,0)$ and $|\Vec{a}_{2}| = \sqrt{y_{0}^2-x_{0}^2-z_{0}^2}/2$,
\begin{equation}
    \Vec{a}_{2}^{(z, x, y)} =U_{-\alpha}\Vec{a}_{2}^{(z', x', y)} = \frac{1}{2}\left(\sqrt{y_{0}^{2}-x_{0}^{2}-z_{0}^{2}} \cos{\alpha}, -\sqrt{y_{0}^{2}-x_{0}^{2}-z_{0}^{2}} \sin{\alpha}, 0\right) \ \ \ .
    \label{eq:alpha2}
\end{equation}
It is useful to express this axis in terms of the $\sigma_2$ CM symplectic eigenvalues $\nu_+ \geq \nu_- \geq 1$ through $y_0^2 - x_0^2 - z_0^2 = \left(\nu_+^2 -1 \right) \left( \nu_-^2-1\right)/\gamma_{c} \geq 0$. Thus, Eqs.~\eqref{eq:twomodeGparam}--\eqref{eq:alpha2} provide a single-variable ($\theta$) parametrization of the optimal X leading to the minimal entanglement GEOF pure state.

Shifting the angular-origin on the ellipse from the vertex closest to $\sigma_\phi$ to the point of maximal $x$-component, $\theta' = \theta - \phi$ with $ \phi = -\arctan \frac{z_0\sqrt{y_{0}^{2}-x_{0}^{2}-z_{0}^{2}}}{x_0 y_0} + \pi\frac{1-\text{sgn}_0\left( x_0\right)}{2}$, the local determinant is expressed as,
\begin{multline}
    m(\theta') = 1 + \Biggl\{  c_{+}\gamma_{c}- c_{-} + \cos{\theta'} \sqrt{\gamma_a \gamma_b} \Biggl\}^2 \times \Biggl\{2\gamma_{c}(a^2+b^2+2c_{-}c_{+}) + 2\gamma_{c}\sin{\theta'}(a^2-b^2)\sqrt{1-\frac{\left( c_{+} \gamma_{c} +c_{-}\right)^2}{\gamma_a \gamma_b}} \\ - 2\gamma_{c}\cos{\theta'}\frac{\left[ 2 a b c_{-}^{3} + (a^2+b^2)c_{+}c_{-}^{2}+\left((1-2b^2)a^2+b^2\right)c_{-}-a b (a^2+b^2-2) c_{+} \right]}{\sqrt{\gamma_a \gamma_b}} \Biggl\}^{-1} \ \ \ ,
    \label{eq:correctversion}
\end{multline}
where $\gamma_a \equiv a \gamma_{c} -b \geq 0$ and $\gamma_b \equiv b \gamma_{c}  -a \geq 0 $ have semi-definite properties arising from the diagonal elements of the physicality condition $\sigma_\phi - \sigma_\pi^{-1} \geq 0$ being principle submatrices of a PSD matrix, and thus PSD as well. Because $ \gamma_a  \gamma_b - \left( c_{+} \gamma_{c} +c_{-}\right)^2 =(\nu_{+}^2-1)(\nu_{-}^2-1)\gamma_{c} \geq 0$, $m(\theta')$ is a well-defined real function when $\gamma_a  \gamma_b \neq 0$. When $\gamma_a \gamma_b = 0$, the physicality condition requires $\sigma_\phi - \sigma_\pi^{-1} =  e|e\rangle \langle e|$ with $|e\rangle = \begin{pmatrix} 0 & 1\end{pmatrix}^T$ or  $|e\rangle = \begin{pmatrix} 1 & 0\end{pmatrix}^T$, such that the $DCV$ is constrained to a one dimensional geometry $\pi/4$ from vertical within a plane of constant $x$. In this case, the minimum and maximum values of the local determinant are given by $X_{opt} = \sigma_{\pi}^{-1}$ and $X_{opt} = \sigma_{\phi}$. Interestingly, such states with physicality conditions are known to have minimum negativity for fixed global and local purities (GLEMS)~\cite{adesso2004determination,It1} and admit purification with the addition of only a single-mode Gaussian system~\cite{It1}.

Updating Eq.~(29) in Ref.~\cite{It3}, Eq.~\eqref{eq:correctversion} allows direct computation of stationary points with $\partial_{\theta'} m=0$. Although minimizing Eq.~\eqref{eq:correctversion} is generally challenging, certain simplifications can be made in this direction. First, observe that the first-order partial derivative, $\partial_{\theta'} m$, results in a fraction. As a squared real function for physical CMs, the denominator of this fraction is always positive. The extrema are thus calculated from roots of the numerator,
\begin{equation}
    \gamma_{c} \underbrace{(N_{1} + N_{2}\cos{\theta'})}_\text{separable roots} \underbrace{(N_{3}\cos{\theta'}+N_{4}\sin{\theta'}+N_{5}\sin^{2} \theta'+N_{6}\cos{\theta'}\sin{\theta'}+N_{5})}_\text{entangled roots} =0 \ \ \ ,
    \label{eq:rooteq}
\end{equation}
where $N_{1}$ to $N_{6}$ are functions involving $a,b,c_{+},c_{-}$. Computational investigations indicate that Eq.~\eqref{eq:rooteq} can have up-to-six physical solutions within a period, with two arising from the second term and four from the third term. For the second term, $N_{1} = c_+ \gamma_c - c_-$ and $N_{2} = \sqrt{\gamma_a  \gamma_b}$. Denote symplectic eigenvalue of PT CM $\Tilde{\nu}_{+} \geq 1$ and $\Tilde{\nu}_{+} \geq \Tilde{\nu}_{-}$. If $\sigma_{2}$ is separable, $\Tilde{\nu}_{-} \geq 1$~\cite{Simonreflection,Duanlocaltrans}, which can be rephrased as constraints on $N_{1}$ and $N_{2}$ as, $N_{2}^2-N_{1}^2=(\Tilde{\nu}_{+}^2-1)(\Tilde{\nu}_{-}^2-1)\gamma_{c} \geq 0$, resulting $1 \geq N_{1}/N_{2} \geq -1$. Therefore, optimal solution for separable $\sigma_{2}$ GEOF reads,
\begin{equation}
    \theta'^{*}=\pm \arccos{\frac{c_{-}-\gamma_{c} c_{+}}{\sqrt{\gamma_a  \gamma_b}}} \ \ \ .
\end{equation}
These solutions arising from the second term are geometrically associated with the ellipse crossing the $x = 0$ plane, a necessary and sufficient condition for two-mode separability. For entangled $\sigma_{2}$, the second term has zero roots, and up-to-four roots of Eq.~\eqref{eq:rooteq} come from the third term. 

In the seminal work on entanglement of formation (EOF)~\cite{akbari2015entanglement}, it is demonstrated that the EOF for an arbitrary two-mode Gaussian state is equivalent to its GEOF, thereby confirming the conjecture proposed in Ref.~\cite{ivan2008entanglementformationgaussianstates} and pursued in Ref.~\cite{marian2008entanglement}. With the closed formula in Eq.~\eqref{eq:correctversion}, a fast and arbitrary-precision determination of EOF of an arbitrary two-mode Gaussian state is given by Eq.~\eqref{eq:rooteq}. While the method present in this section offers advantage over methods involving coupled equations~\cite{ivan2008entanglementformationgaussianstates,marian2008entanglement,akbari2015entanglement}, as well as single-parameter searching~\cite{WolfGEOF} and optimization~\cite{2019geof}, it is also expected to contribute toward analytical bounds of EOF for arbitrary two-mode Gaussian state as investigated in Refs.~\cite{PhysRevA.96.062338,2019geof}.

\begin{figure}[t!]
\centering
\includegraphics[width=0.4\textwidth]{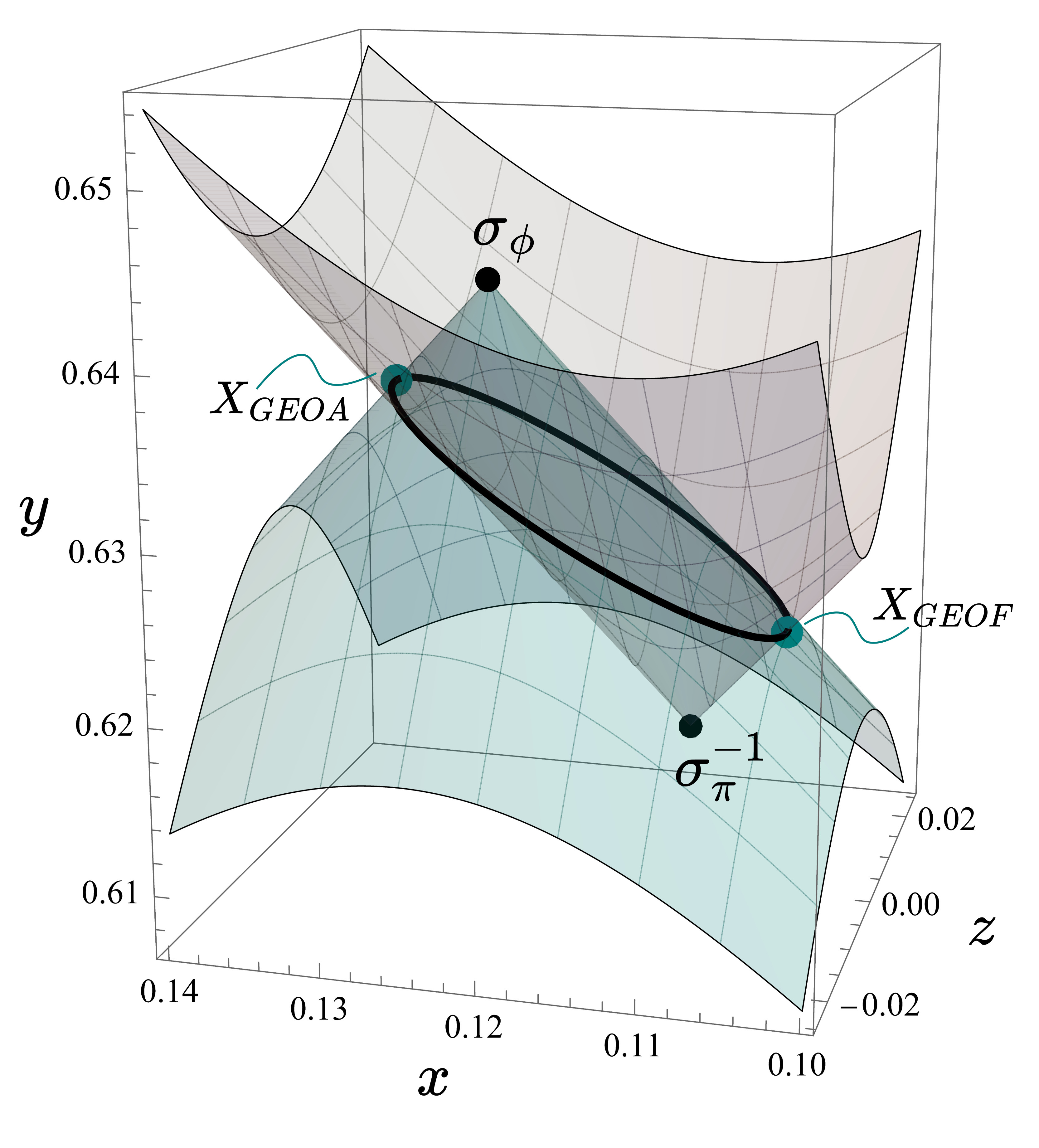}
\caption{$DCV$ for symmetric two-mode Gaussian states with $d=1$, $\tilde{r}=0$ and $m=1$. The vertices of the $DCV$ and the rim are shown in black. The solutions for GEOA and GEOF are shown in teal, which are symmetrically distributed at either end of the major axis on the elliptical rim.}
\label{fig:coneconediagram1}
\end{figure}

For symmetric Gaussian states, the optimal solutions to $X_{GEOF}$ and $X_{GEOA}$ reside at opposing points on the major axis of the $DCV$ rim, as derived in Appendix~\ref{sec:appbsec1}. This is depicted in Fig.~\ref{fig:coneconediagram1} for two neighboring lattice sites within a free scalar field vacuum  $(d, \tilde{r}, m) = (1,0,1)$. While the corresponding CM is not in the form of Eq.~\eqref{eq:twomodeGparam}, the underlying pure state resulting in GEOF and GEOA can still be contextualized on the major axis within the $DCV$ framework (see end of Appendix~\ref{sec:appbsec1}) to derive that $X_{GEOF}$ and $X_{GEOA}$ resides on the elliptical rim of intersecting positivity cones.

Previous calculations of the GEOA for tensor-product two-mode Gaussian states and GLEMS, together with the current results for symmetric two-mode Gaussian states, consistently yield an underlying pure state that lies on the boundary of the DCV. This observation, together with additional numerical studies, leads to the conjecture that such a property holds for the GEOA of arbitrary two-mode Gaussian states.

\subsection{Multimode \texorpdfstring{$DCV$}{DCV}s in higher-dimensions and inter-\texorpdfstring{$DCV$}{DCV} distance}
\label{sec:appcsec2}

In the context of entanglement identification, this section creates a conic framework that extends to the multimode regime. For Gaussian states sharing structural properties with the scalar field vacuum~\cite{gao2024detectingspacelikevacuumentanglement}, a pair of $DCV$s are defined below, where their separation serves as a necessary and sufficient quantifier of entanglement.

In Ref.~\cite{Gaussianboundent}, it is shown that a CM is separable if and only if there exists two CMs $\sigma_{A}$ and $\sigma_{B}$ such that $\sigma \geq \sigma_{A} \oplus \sigma_{B}$. In Ref.~\cite{WolfGEOF}, it is shown that $\phi\pi$-uncorrelated CMs always have a valid pure state decomposition of minimal entanglement that is also $\phi\pi$-uncorrelated, i.e., $Y = 0$ in Eq.~\eqref{eq:pureCMform}. Because parity symmetry can be simultaneously maintained within these noise identification families, a separable, $\phi\pi$-uncorrelated, AB-symmetric state admits a pure state decomposition of the form,
\begin{equation}
    X =
\begin{pmatrix}
    X_{A} & X_{AB} \\
    X_{AB} & X_{A} \\
\end{pmatrix} , \quad  X_{AB} = 0, \quad Y = 0 \ \ \ .
\end{equation}
Therefore, there exists $X_{A}$ and $X_{AB} = 0$ such that,
\begin{equation}
\scalemath{1}{\begin{cases}
      \sigma_{\phi,A} + \sigma_{\phi,AB} \geq X_{A}+X_{AB} \geq  \left(\sigma_{\pi}^{-1}\right)_{A} + \left(\sigma_{\pi}^{-1}\right)_{AB} \\
      \sigma_{\phi,A} - \sigma_{\phi,AB} \geq X_{A}-X_{AB} \geq  \left(\sigma_{\pi}^{-1}\right)_{A} - \left(\sigma_{\pi}^{-1}\right)_{AB}
    \end{cases}} ,
    \label{eq:twoDCVs1}
\end{equation}
if and only if the corresponding AB symmetric and $\phi\pi$-uncorrelated Gaussian state is separable. Equation~\eqref{eq:twoDCVs1} generalizes the $DCV$ framework from a three dimensional vector space to higher dimensions~\footnote{Note Eq.~\eqref{eq:twoDCVs1} can be alternatively derived from Gaussian Rényi-2 (GR2) entanglement of formation, which relates to phase-space sampling entropy~\cite{adessoGR2}. With Eq.~\eqref{eq:pureCMform}, the local determinant of the underlying pure CM for disjoint scalar field vacuum may be written as,
\begin{equation}
\det \left( \left(\sigma_{0}\right)_{A} \right)= \det \left(X_{A}\right)\times \\  \det \left(\left(X^{-1}\right)_{A}+Y_{AB} \left(X/X_{A} \right)Y_{AB}^{T}\right) \geq \det \left( \left(\sigma_{0}\right)_{A} \left( Y =0\right)\right) \ \ \ ,
\label{eq:localdet}
\end{equation}
where $X/X_{A}$ denotes the Schur complement. In the $\phi\pi$-uncorrelated multimode context, an underlying pure state CM with $Y=0$ has previously been shown to be always available~\cite{WolfGEOF}. Therefore, Eq.~\eqref{eq:localdet} simplifies GR2 EOF to,
\begin{equation}
\min_{X}\{ m\left(X\right) \quad | \quad   \sigma_{\pi}^{-1} \leq X \leq \sigma_{\phi} \}, \quad m\left(X\right)=    \frac{\det\left(X_{A}\right)\det\left(X_{B}\right)}{\det\left(X\right)} \\ = \frac{\det\left(\left(X^{-1}\right)_{A}\right)\det\left(\left(X^{-1}\right)_{B}\right)}{\det\left(X^{-1}\right)} \ \ \ ,
\label{eq:localdet3}
\end{equation}
which reproduces Eq.~\eqref{eq:twoDCVs1} for a separable pure state decomposition. Furthermore, from Jacobi's formula~\cite{hornmatrixana}, it is seen that perturbing $X$ by a block diagonal PSD matrix will result in negative first derivative. Therefore, sending $X < \sigma_{\phi}$ to $X \leq \sigma_{\phi}$ via the chosen perturbation decreases the cost function, and a similar argument applies to $X^{-1} < \sigma_{\pi}$. This directly generalizes the constrained optimization in Appendix~\ref{sec:appcsec1} to multimode Gaussian states.}, where $DCV_{+}$ and $DCV_{-}$ are defined according to the first and second matrix inequalities. Distance between these two $DCV$s in the matrix vector space can be quantified by the matrix Frobenius norm, which reads,
\begin{equation}
    \xi \equiv || X_{AB} ||=\sqrt{\Tr \left[ X_{AB}^{2} \right]} \ \ \ .
\label{eq:d0trxab}   
\end{equation}
The necessary and sufficient separability condition can thus be expressed as the existence of $\xi=0$, i.e., the intersection between $DCV_{+}$ and $DCV_{-}$.

\begin{figure}[t!]
\centering
\includegraphics[width=0.45\textwidth]{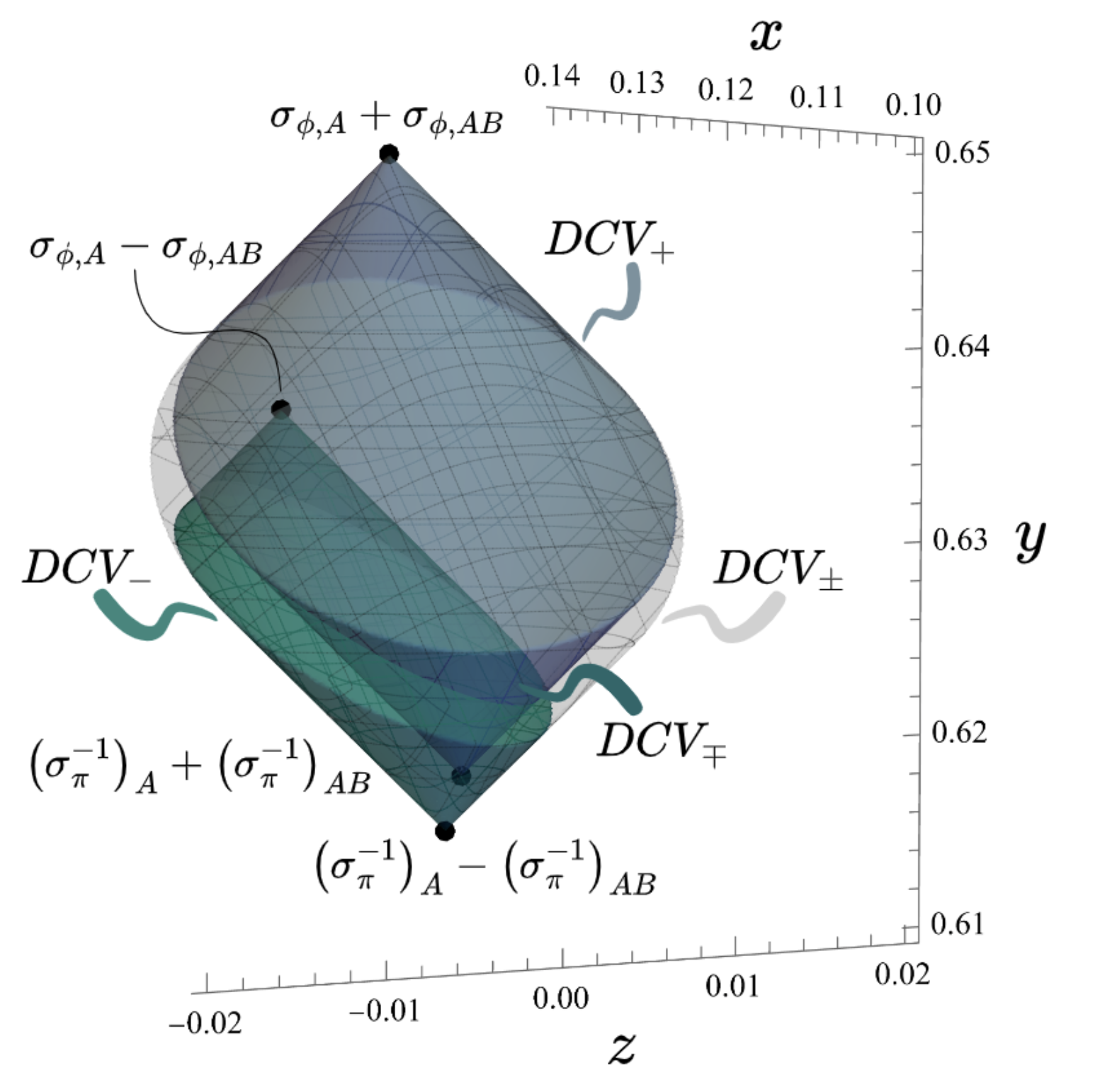}
\caption{Multimode $DCV$s for spacelike regions of the lattice scalar field vacuum, $(d, \tilde{r}, m) = (2,2,1)$, where $DCV_{\pm} \supseteq DCV_+ \cup DCV_-$. As a $(2_A \times 2_B)$ separable state, the matrix inequalities discussed in the text indicate that the lower vertex of $DCV_+$, $\left(\sigma_{\pi}^{-1}\right)_{A} + \left(\sigma_{\pi}^{-1}\right)_{AB}$, resides inside $DCV_-$ and the upper vertex of $DCV_-$, $\sigma_{\phi,A} - \sigma_{\phi,AB}$, resides inside $DCV_+$. Thus $\xi$ = 0 and $DCV_{\mp} = DCV_+ \cap DCV_-$ has a non-zero volume.}
\label{fig:coneconediagram2}
\end{figure}

In a previous work~\cite{gao2024detectingspacelikevacuumentanglement}, the Peres-Horodecki criterion~\cite{peresoriginalN,HORODECKIoriginalN,Simonreflection} is proven to provide a necessary and sufficient condition for determining whether disjoint regions in the free scalar field vacuum are separable. Reproducing from Ref.~\cite{gao2024detectingspacelikevacuumentanglement}, the criterion in this context of $AB$-symmetric and $\phi\pi$-uncorrelated CMs corresponds to the conditions, 
\begin{equation}
\begin{cases}
      \sigma_{\phi,A} + \sigma_{\phi,AB}  \geq \left(\sigma_{\pi}^{-1}\right)_{A} - \left(\sigma_{\pi}^{-1}\right)_{AB} \\
      \sigma_{\phi,A} - \sigma_{\phi,AB}  \geq \left(\sigma_{\pi}^{-1}\right)_{A} + \left(\sigma_{\pi}^{-1}\right)_{AB}
    \end{cases} \ \ \ .
\label{eq:ncontribute}
\end{equation}
Note that, for a general quantum state, Eq.~\eqref{eq:ncontribute} provides only a necessary condition for the existence of intersection between $DCV_{+}$ and $DCV_{-}$ with $X_{AB}=0$, which is consistent with the Peres-Horodecki criterion generally providing only a necessary condition for separability.

The scalar field vacuum CM in addition satisfies~\cite{gao2024detectingspacelikevacuumentanglement},
\begin{equation}
 \sigma_{\phi,AB} \geq 0, \quad \left(\sigma_{\pi}^{-1}\right)_{AB} \geq 0 \ \ \ ,
 \label{eq:fieldproperty}
\end{equation}
which makes the matrices associated with each of the two $DCV_+$ vertices in Eq.~\eqref{eq:twoDCVs1} greater than or equal to their $DCV_-$ counterparts, respectively. The conditions in Eqs.~\eqref{eq:twoDCVs1} and~\eqref{eq:ncontribute} can thus be expressed as a single hierarchy of the upper and lower $DCV$ vertices:  $DCV_+^{(u)} \geq DCV_-^{(u)} \geq DCV_+^{(\ell)} \geq DCV_-^{(\ell)}$. From the principal submatrix of the physicality condition $\sigma_\phi - \sigma_\pi^{-1} \geq 0$, it is seen that $\sigma_{\phi,A} - \left(\sigma_{\pi}^{-1}\right)_{A} \geq 0$. The first inequality of Eq.~\eqref{eq:ncontribute} is thus automatically satisfied, and the second inequality leads to an AB symmetric and $\phi\pi$-uncorrelated separable state decomposition through the Gaussian classical mixing channel when the negativity vanishes~\cite{gao2024detectingspacelikevacuumentanglement}. When the CM is separable, any $X_{A}$ with $X_{AB}=0$ in the separable $DCV_{\mp} = DCV_- \cap DCV_+$ defined by $\sigma_{\phi,A} - \sigma_{\phi,AB} \geq X_{A} \geq \left(\sigma_{\pi}^{-1}\right)_{A} + \left(\sigma_{\pi}^{-1}\right)_{AB}$, will satisfy Eq.~\eqref{eq:twoDCVs1}. As depicted in Fig.~\ref{fig:coneconediagram2}, the $DCV_{+}$ and $DCV_{-}$ are furthermore included in the encompassing $DCV_{\pm}$ with vertices corresponding to $\sigma_{\phi,A} + \sigma_{\phi,AB}$ and $\left(\sigma_{\pi}^{-1}\right)_{A} - \left(\sigma_{\pi}^{-1}\right)_{AB}$, which leads to an analytical description for maximizing distance $\xi$ between $DCV_+$ and $DCV_-$ in the following section.

\subsection{Maximizing inter-\texorpdfstring{$DCV$}{DCV} distance and \texorpdfstring{$\mathcal{V_{\pm}}$}{Vpm}-SOL MNF pure state identification}
\label{sec:appcsec3}

In this section, it is shown that the pure state identified by $\mathcal{V}_{\pm}$-SOL MNF (Appendix~\ref{sec:appbsec3}) maximizes the inter-$DCV$ distance, $\xi$.  To do so,  a solution saturating the upper bound of a relaxed maximization problem is shown to also be a solution of the original distance maximization.

Consider a relaxation in the distance maximization program,
\begin{equation} 
\scalemath{1}{\begin{aligned} 
&\text{maximize} && \xi^{2} \equiv \Tr \left[ X_{AB}^{2} \right]  \\ &\text{subject to} && \begin{cases}
      X_{A}+X_{AB} \in \text{$DCV_{\pm}$} \\
      X_{A}-X_{AB} \in \text{$DCV_{\pm}$}
    \end{cases}
\end{aligned} \ \ \ ,
\label{eq:ddd}}
\end{equation}
utilizing $DCV_+ \subseteq DCV_{\pm}$ and $DCV_- \subseteq DCV_{\pm}$. Through matrix inequalities,
\begin{equation}
    \begin{cases}
         \scalemath{1}{  \sigma_{\phi,A} + \sigma_{\phi,AB} - X_{A} \pm X_{AB} \geq 0 } \\
      \scalemath{1}{  X_{A} \pm X_{AB} - (\sigma_{\pi}^{-1})_{A}  + (\sigma_{\pi}^{-1})_{AB} \geq 0} \\
    \end{cases} \Rightarrow \quad \sigma_{\phi,A} + \sigma_{\phi,AB}-\left(\sigma_{\pi}^{-1}\right)_{A} + \left(\sigma_{\pi}^{-1}\right)_{AB} \pm 2X_{AB}  \geq 0 \ \ \ ,
\end{equation}
the cost function is observed to have an upper bound,
\begin{equation}
    \scalemath{1}{||\sigma_{\phi,A} + \sigma_{\phi,AB}-\left(\sigma_{\pi}^{-1}\right)_{A} + \left(\sigma_{\pi}^{-1}\right)_{AB}||^{2} \geq 4\xi^{2}} \ \ \ .
\end{equation}
With the choice,
\begin{equation}
\begin{cases}
      \scalemath{1}{ 2X_{A} = \sigma_{\phi,A} + \sigma_{\phi,AB} + (\sigma_{\pi}^{-1})_{A}  - (\sigma_{\pi}^{-1})_{AB} } \\
      \scalemath{1}{2X_{AB} =  \sigma_{\phi,A} + \sigma_{\phi,AB} - (\sigma_{\pi}^{-1})_{A}  + (\sigma_{\pi}^{-1})_{AB} } \\
    \end{cases}  ,
\label{eq:MNFpurestate}
\end{equation}
this upper bound is saturated. Because this solution corresponds to the sum(difference) of $X_A$ and $X_{AB}$ located at the upper(lower) vertex of $DCV_{\pm}$, it is also a solution to the original maximization problem. As discussed in Sec.~\ref{sec:IIB} and formulated in Appendix~\ref{sec:appbsec1}, the underlying pure state identified for GEOA of symmetric two mode Gaussian states is consistent with Eq.~\eqref{eq:MNFpurestate}.

To show that a $\mathcal{V}_{\pm}$-SOL MNF pure state (noise isolated to $\mathcal{V}_{\mp}$) is given by Eq.~\eqref{eq:MNFpurestate}, note that the relative structure of $X_{A} \pm X_{AB}$ embedded within the conic topology remains invariant under $AB$-symmetric and $\phi\pi$-uncorrelated local symplectic transformations, which updates Eq.~\eqref{eq:twoDCVs1} through linear transformation $L$ (as introduced in Eq.~\eqref{eq:design1} for transformations in $C$). Therefore, this property of Eq.~\eqref{eq:MNFpurestate} is free to be shown through a strategic choice of the locally consolidated basis obtained by applying a modified version of Eq.~\eqref{eq:consopmsol}~\footnote{Since structural invariance is preserved only under $AB$-symmetric and $\phi\pi$-uncorrelated local symplectic transformations, the minus sign in Eq.~\eqref{eq:consopmsol} is discarded. This will simply produce post-consolidation $\tilde{S}$ row vectors in $\mathcal{V}_{\pm}$ that are $AB$-symmetric.}, $\sigma' \stackrel{\text{reorder}}{=} \left( \tilde{S}_{\pm} \oplus \tilde{S}_{\pm} \right) \sigma \left( \tilde{S}_{\pm} \oplus \tilde{S}_{\pm} \right)^{T}$. In this basis it follows that~\footnote{Note that any matrix of the form $M \equiv \begin{pmatrix}
M_{A} & M_{AB} \\
M_{AB} & M_{A}
\end{pmatrix}$ has a relation $(M_{A} \pm M_{AB})^{-1} = (M^{-1})_{A} \pm (M^{-1})_{AB}$ when relevant inverses are well defined.},
\begin{equation}
     \left(\sigma'\right)_{\phi,A} + \left(\sigma'\right)_{\phi,AB}  = \left( \left(\left(\sigma'\right)_{\pi}^{-1}\right)_{A}  - \left(\left(\sigma'\right)_{\pi}^{-1}\right)_{AB} \right)^{-1} \equiv \Tilde{D}_{\pm,half} \ \ \ ,
\end{equation}
which can be seen by noticing that Eq.~\eqref{eq:consopmsol} diagonalizes $\Tilde{\sigma}_{\pm}$ of Eq.~\eqref{eq:pmmp}. Here, $\Tilde{D}_{\pm,half} \geq \mathbb{I}$~\cite{gao2024detectingspacelikevacuumentanglement} is a $d$-dimensional diagonal matrix with the non-degenerate symplectic eigenvalues of $\Tilde{\sigma}_{\pm}$ on the diagonal. As such, Eq.~\eqref{eq:MNFpurestate} becomes,
\begin{equation}
\begin{cases}
      \scalemath{1}{ 2\left(X'\right)_{A} = (\Tilde{D}_{\pm,half} + \Tilde{D}_{\pm,half}^{-1}) } \\
      \scalemath{1}{2\left(X'\right)_{AB} = (\Tilde{D}_{\pm,half} - \Tilde{D}_{\pm,half}^{-1}) } \\
    \end{cases}  \ \ \ .
\label{eq:MNFpurestate1}
\end{equation}
The relationships $\left(X'\right)_{A}^{2}-\left(X'\right)_{AB}^{2} = \mathbb{I}$ and $\left(X'\right)_{AB}\geq 0$ allow parametrization of $\left(X'\right)_{A}$ with $\cosh$ and $\left(X'\right)_{AB}$ with $\sinh$, realizing the underlying pure state as a tensor product of positively squeezed TMSVS. Since the symplectic eigenvalues of $\tilde{\sigma}_{\pm}$ are equal to those of the block-combined pure state governed by Eq.~\eqref{eq:MNFpurestate1}, the noise is isolated to $\mathcal{V}_{\mp}$. Therefore, the pure state of Eq.~\eqref{eq:MNFpurestate} is locally equivalent to the $\mathcal{V}_{\pm}$-SOL MNF pure state.

\subsection{Minimizing inter-\texorpdfstring{$DCV$}{DCV} distance and assisted measurement profiles}
\label{sec:appcsec4}

After introducing semidefinite programming with quadratic cost function, this section presents a technique for upper-bounding many-body GEOF guided by minimization of distance $\xi$ between the $DCV$ pair introduced above. Demonstrating its value by application in the scalar field vacuum, this approach enables the first conclusive identification of exponentially decaying underlying pure-state entanglement between disjoint massless field regions. As discussed in Sec.~\ref{sec:IIIB}, this result is the lowest known upper bound to the spacelike GEOF in the field vacuum, and the first that is parametrically tight to the exponential decay of the available negativity.

Semidefinite programming (SDP)~\cite{vandenberghe1996semidefinite} is a class of convex optimization problems~\cite{boyd2004convexschurcomp} that generalizes linear programming~\cite{dantzig2002linear} and has broad application, e.g., in combinatorial optimization~\cite{alizadeh1995interior,benson2000solving}, signal processing~\cite{meng2008semidefinite,huang2009rank}, and quantum information science~\cite{skrzypczyk2023semidefinite,mironowicz2024semi}. An SDP is typically formulated as,
\begin{equation} 
\begin{aligned} 
&\text{minimize} && \text{Tr}(A Q) \\ &\text{subject to} && Q \geq 0, \quad \text{Tr}(B_i Q) \leq b_i, \quad \forall i 
\end{aligned} \ \ \ ,
\label{eq:d1}
\end{equation} 
where $Q$ is a symmetric matrix variable, $A$ and $B_i$ are given symmetric matrices, and $b_i$ are given constants. This structure makes SDP a special case of convex optimization~\cite{boyd2004convexschurcomp} where the decision variable is a matrix.

Before presenting the application to Gaussian states, it is advantageous to introduce epigraph transformation~\cite{boyd2004convexschurcomp}, which incorporates quadratic cost functions into the SDP framework. To see this, consider the equivalence,
\begin{equation} 
\begin{aligned} 
&\text{minimize} && \boldsymbol{x}^{T} N \boldsymbol{x}  \quad \Longleftrightarrow \quad \begin{aligned} 
&\text{minimize} && t \\ &\text{subject to} && \boldsymbol{x}^{T} N \boldsymbol{x}   \leq t 
\end{aligned} 
\end{aligned} \ \ \ ,
\label{eq:d2}
\end{equation} 
where $N = N^T \geq 0$, $\mathbf{x}$, and $t \geq 0$ are a real matrix(given), vector, and scalar, respectively. Note that $N$ admits an eigendecomposition, $N = O^{T}DO= R^{T} R$, with $O$ an orthogonal matrix and $R\equiv \sqrt{D}O$. The quadratic constraint can therefore be rephrased as,
\begin{equation}
    t   - \boldsymbol{x}^{T} R^{T} R \boldsymbol{x} \geq 0 \quad \Longleftrightarrow \quad \begin{pmatrix} \mathbb{I} & R \boldsymbol{x} \\ \boldsymbol{x}^{T} R^{T} & t \end{pmatrix} \geq 0  \ \ \ ,
    \label{eq:d3}
\end{equation}
where the equivalence can be seen by expressing the semidefinite properties of the matrix using the Schur complement. Together, Eqs.~\eqref{eq:d2} and~\eqref{eq:d3} allow the use of quadratic cost functions within the framework of SDP.

\begin{figure}[t!]
\centering
\includegraphics[width=1\textwidth]{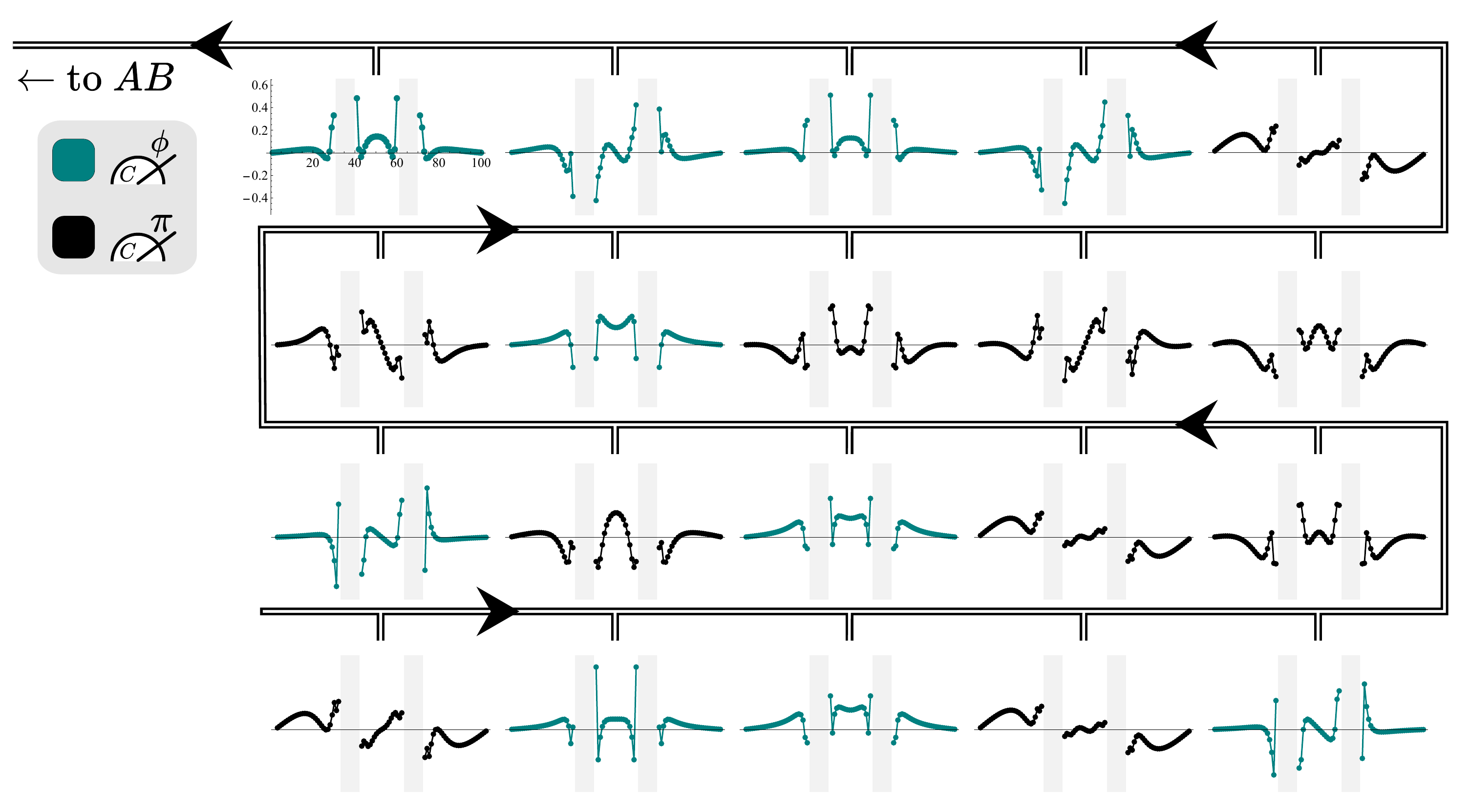}
\caption{Collective profiles in $\pi$ and $\phi$ spaces for assisted projective measurement obtained from the eigendecomposition of the noise minimizing distance $\xi$ between two multimode $DCV$s. The system configuration and measurement/communication procedure are the same as that described in Fig.~\ref{fig:fig4} of Appendix~\ref{sec:appbsec3}. In the order connected to the classical communication line, $\pi$ and $\phi$ profiles correspond to decreasing $Y$ eigenvalue. A symplectic transformation, $S_{\{ \langle p| \} }$, localizing these profiles within the volume $C$ prior to measurement may be constructed following Eqs.~\eqref{eq:gsreplace1} and~\eqref{eq:gsreplace2}.}
\label{fig:fig8}
\end{figure}

Following Appendixes~\ref{sec:appcsec2} and~\ref{sec:appcsec3}, inter-$DCV$ distance can be expressed by Eq.~\eqref{eq:d0trxab}, i.e., $\xi \equiv ||X_{AB}||$. With symmetric matrix variables $P = X_{A} + X_{AB}$, $M = X_{A} - X_{AB}$, and denoting $\boldsymbol{p_{0}}$, $\boldsymbol{m_{0}}$ as a vectorization of the upper triangle of $P$, $M$,
\begin{equation}
 \xi^{2}=  \boldsymbol{\begin{pmatrix} p_{0} \\ m_{0} \end{pmatrix}}^{T} \begin{pmatrix} d_{0} & -d_{0} \\ -d_{0} & d_{0} \end{pmatrix} \boldsymbol{\begin{pmatrix} p_{0} \\ m_{0} \end{pmatrix}}, \quad \begin{cases}
      \boldsymbol{v_{0}} \equiv (\underbrace{v_{11},v_{12},...,v_{1d}}_\text{d},\underbrace{v_{22},v_{23}...,v_{2d}}_\text{d-1},...,v_{dd})  \\
     d_{0} \equiv \text{diag}(1,\underbrace{2,...,2}_\text{d-1},1,\underbrace{2,...,2}_\text{d-2},...,1,\underbrace{2}_\text{1},1)/4 \\
    \end{cases} \ \ \ ,
\end{equation}
where $ \boldsymbol{v_{0}} \in \{  \boldsymbol{p_{0}}, \boldsymbol{m_{0}} \}$. Since $d_{0}>0$, there exists $R_{0}^{T}R_{0}=\begin{pmatrix} d_{0} & -d_{0} \\ -d_{0} & d_{0} \end{pmatrix}$, such that the distance metric between the $DCV$ pair can be minimized with an SDP of the form,
\begin{equation}
    \scalemath{0.69}{\begin{aligned} 
&\text{minimize} && t \\ &\text{subject to} && \begin{pmatrix} \mathbb{I} & R_{0} \boldsymbol{\begin{pmatrix} p_{0}  \\ m_{0} \end{pmatrix}}& 0 \\ \boldsymbol{\begin{pmatrix} p_{0} \\ m_{0} \end{pmatrix}}^{T} R_{0}^{T} & t  & 0 \\ 0 & 0& \left( \sigma_{\phi,A} + \sigma_{\phi,AB} - P \right) \oplus \left( P-\left(\sigma_{\pi}^{-1}\right)_{A} - \left(\sigma_{\pi}^{-1}\right)_{AB} \right) \oplus \left( \sigma_{\phi,A} - \sigma_{\phi,AB} - M \right) \oplus \left( M -\left(\sigma_{\pi}^{-1}\right)_{A} + \left(\sigma_{\pi}^{-1}\right)_{AB} \right) \\ \end{pmatrix} \geq 0 
\end{aligned} }  ,
\label{eq:SDPscalar}
\end{equation}
where $R_{0}$ and $\sigma$ are given, and the lower four blocks ensure that $P \in DCV_+$ and $M \in DCV_-$. Note that this optimization program can be rephrased in the framework of Eq.~\eqref{eq:d1} by identifying the projector $A = |d(d+1)\rangle \langle d(d+1)|$ such that $\text{Tr}\left(AQ\right) = t$, where $Q$ is the PSD matrix in Eq.~\eqref{eq:SDPscalar}. Numerically evaluating this SDP program with the arbitrary-precision, parallelized semidefinite program solver SDPB~\cite{simmons2015semidefinite,Landry:2019qug} leads to the new upper bound for GEOF presented in Fig.~\ref{fig:plot}. 

It is interesting to observe the span of underlying pure-state negativities identified by the $DCV$ distance minimization (Eq.~\eqref{eq:SDPscalar}) and maximization (Appendix~\ref{sec:appbsec3} as connected in Appendix~\ref{sec:appcsec3}). Continuing with the 100-site scalar field example with $\left( m, d, \tilde{r}\right) = \left(0, 10, 20 \right)$, minimizing this distance for an upper-bound of GEOF leads to $\mathcal{N}_p = 0.00173$, significantly less than the distance maximization of $\mathcal{N}_p = 1.21$ found above to lower-bound GEOA. Figure~\ref{fig:plot} indicates that these two physical quantities are on exponentially distinct trajectories as a function of $\tilde{r}$ separation. In the same way that Fig.~\ref{fig:fig4} used the techniques of Appendix~\ref{sec:appa} to calculate $C$-measurement profiles corresponding to the $\mathcal{V}_{\pm}$-SOL MNF rank-one decomposition of the noise whose removal reveals the $\xi$-maximized pure state, Fig.~\ref{fig:fig8} shows $C$-measurement profiles that would reveal the $\xi$-minimized $AB$ pure state. Here, the collective modes for volume measurement can again be designed solely from the $2d$ volume normal modes that are entangled with the patches, since the resulting $\xi$-minimized noise has rank $2d$. These profiles, governed by a rank-one decomposition of the noise, can be structured to provide a systematic convergence in the negativity and purity. For example, even without a specialized rank-one decomposition protocol, e.g., comparable to that provided by MNF above, the set of measurements corresponding simply to decreasing order of the eigenvalues produces a sequence of deviations from unit purity: $0.583, 0.461, 0.415, 0.371, 0.283, 0.185, 0.103, 0.0195, 0.0112, 0.00378, 0.00267, 0.00108, 0.000534, 1.93 \times 10^{-6}, 8.78 \times 10^{-7}, 4.75 \times 10^{-10}, 3.02 \times 10^{-10}, 1.73 \times 10^{-14}, 3.53 \times 10^{-15}, 0$. As such, a truncation on the number of measurements may be designed to coincide with experimental capabilities. 

A joint analysis of Fig.~\ref{fig:plot}, Fig.~\ref{fig:fig4}, and Fig.~\ref{fig:fig8} reveals that entanglement purified between $A$ and $B$, i.e., via classical communication of collective measurements performed in $C$, can vary exponentially depending upon the choice of measurement basis in $C$. While the measurement profiles may appear similar, the resulting purified entanglement differs significantly. This is further evidence that entanglement is highly sensitive to measurement profiles, consistent with the observations of Ref.~\cite{gao2024detectingspacelikevacuumentanglement} in the context of $AB$ profiles for entanglement detection.

\end{document}